\documentclass[sigconf, screen, anonymous=false, authordraft=False]{acmart}

\usepackage{subcaption}
\usepackage{multirow}
\usepackage{tabularx}
\usepackage{dcolumn} 
\newcolumntype{d}[1]{D{.}{.}{#1}}
\usepackage{gensymb}
\usepackage{soul}

\AtBeginDocument{%
  }

\copyrightyear{2026}
\acmYear{2026}
\setcopyright{cc}
\setcctype{by-nc-nd}
\acmConference[CHI '26]{Proceedings of the 2026 CHI Conference on Human Factors in Computing Systems}{April 13--17, 2026}{Barcelona, Spain}
\acmBooktitle{Proceedings of the 2026 CHI Conference on Human Factors in Computing Systems (CHI '26), April 13--17, 2026, Barcelona, Spain}
\acmPrice{}
\acmDOI{10.1145/3772318.3790313}
\acmISBN{979-8-4007-2278-3/2026/04}

\begin{document}

\title{Understanding the Effects of Interaction on Emotional Experiences in VR}

\settopmatter{authorsperrow=3}

\author{Zheyuan Kuang}
\orcid{0009-0009-0184-6159}
\affiliation{%
 \institution{The University of Sydney}
 \streetaddress{Camperdown/Darlington}
 \city{Sydney}
 \country{Australia}}
\email{zheyuan.kuang@sydney.edu.au}

\author{Tinghui Li}
\orcid{0009-0009-3121-8997}
\affiliation{%
 \institution{The University of Sydney}
 \streetaddress{Camperdown/Darlington}
 \city{Sydney}
 \country{Australia}}
\email{tinghui.li@sydney.edu.au}

\author{Weiwei Jiang}
\orcid{0000-0003-4413-2497}
\affiliation{%
 \institution{Nanjing University of Information Science and Technology}
 \city{Nanjing}
 \country{China}}
\email{weiweijiangcn@gmail.com}

\author{Sven Mayer}
\orcid{0000-0001-5462-8782}
\affiliation{%
  \institution{TU Dortmund University}
  \city{Dortmund}
  \country{Germany}
 }
\affiliation{%
  \institution{Research Center Trustworthy Data Science and Security}
  \city{Dortmund}
  \country{Germany}}
\email{info@sven-mayer.com}

\author{Flora Salim}
\orcid{0000-0002-1237-1664}
\affiliation{%
 \institution{University of New South Wales}
 \streetaddress{Kensington}
 \city{Sydney}
 \country{Australia}}
\email{flora.salim@unsw.edu.au}

\author{Benjamin Tag}
\orcid{0000-0002-7831-2632}
\affiliation{%
 \department{School of Computer Science and Engineering}
 \institution{University of New South Wales}
 \streetaddress{Kensington}
 \city{Sydney}
 \state{New South Wales}
 \country{Australia}}
\email{benjamin.tag@unsw.edu.au}

\author{Anusha Withana}
\orcid{0000-0001-6587-1278}
\affiliation{%
 \department{School of Computer Science}
 \institution{The University of Sydney}
 \streetaddress{Camperdown/Darlington}
 \city{Sydney}
 \state{NSW}
 \country{Australia}}
\email{anusha.withana@sydney.edu.au}

\author{Zhanna Sarsenbayeva}
\orcid{0000-0002-1247-6036}
\affiliation{%
 \institution{The University of Sydney}
 \streetaddress{Camperdown/Darlington}
 \city{Sydney}
 \country{Australia}}
\email{zhanna.sarsenbayeva@sydney.edu.au}

\renewcommand{\shortauthors}{Kuang et al.}

\begin{abstract}
Virtual reality has been effectively used for eliciting emotions, yet most research focuses on the intensity of affective responses rather than on how interaction influences those experiences. To address this gap, we advance a validated VR emotion-elicitation dataset through two key extensions. First, we add a new high-arousal, high-valence scene and validate its effectiveness in a within-subject study (N=24). Second, we incorporate interactive elements into each scene, creating both interactive and non-interactive versions to examine the impact of interaction on emotional responses. We evaluate interaction through a multimodal approach combining subjective ratings and physiological signals to capture both conscious and unconscious affective responses. Our evaluation study (N=84) shows that interaction not only amplifies emotions but modulates them in context, supporting coping in negative scenes and enhancing enjoyment in positive scenes. These findings highlight the potential of scene-tailored interaction for different applications, where regulating emotions is as important as eliciting them.
\end{abstract}




\begin{teaserfigure}
 \centerline{\includegraphics[width=0.845\linewidth]{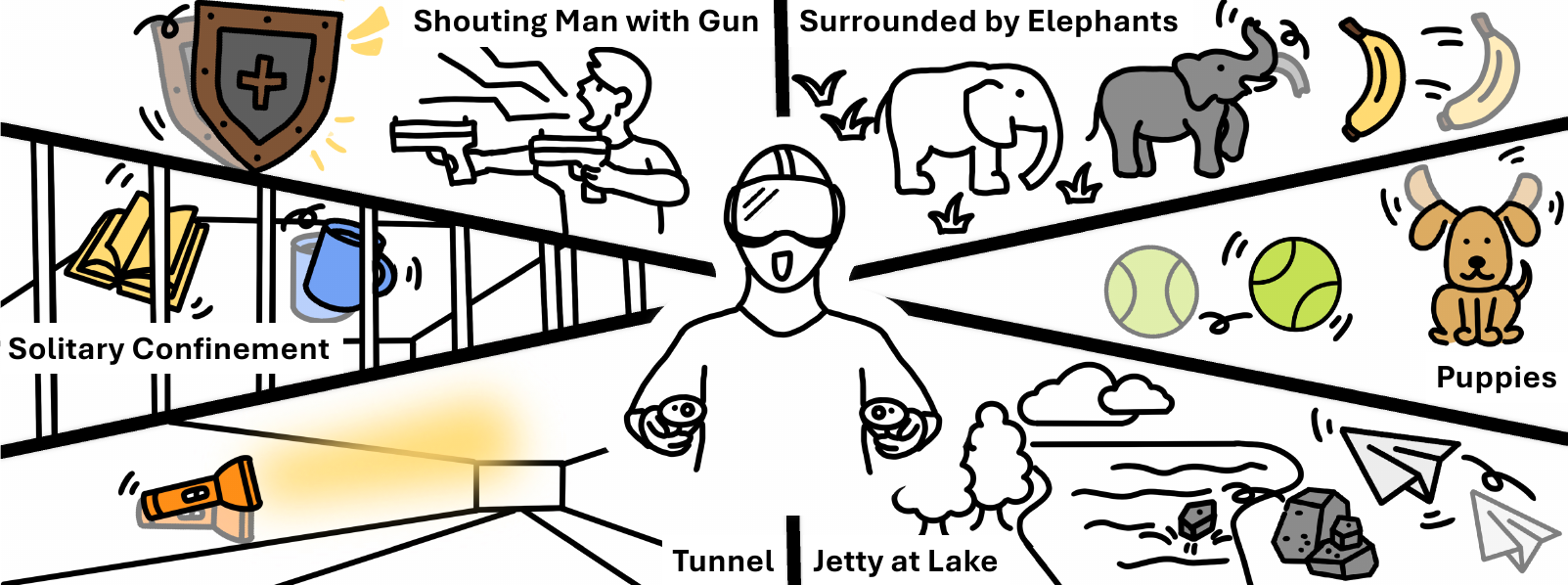}}
 \caption{Illustration of the participant under six virtual reality scenes that could elicit emotions with interactive objects highlighted in color. }
 \Description{A person wearing a VR headset stands in the middle holding two controllers, while six wedge shaped panels around them depict the study scenes: Solitary Confinement, Tunnel, Jetty at Lake, Puppies, Surrounded by Elephants, and Shouting Man with Gun. Each panel contains simple line drawings of key scene elements (for example, a corridor like tunnel, a lakeside jetty with trees and rocks, puppies with tennis balls, two elephants, and a man aiming a gun). Objects intended for interaction are highlighted using color, including items such as a flashlight casting a yellow beam in the tunnel, green tennis balls, yellow bananas, and other colored props inside the scene panels.}
 \label{fig:teaser}
\end{teaserfigure}

\maketitle


\section{Introduction}
Emotions play a central role in shaping human experience, influencing perception, decision-making, and behaviour~\cite{lazarus1991emotion}. Virtual reality (VR) has emerged as a powerful tool for inducing emotions, offering immersive and ecologically valid environments where affective states can be readily elicited for integrated emotion experiences~\cite{jiang2024immersive}. Prior research has successfully used immersive 360$\degree$ videos, photorealistic environments, and virtual environments to evoke a range of emotions~\cite{li2017public, schone2023library, jiang2024immersive}. However, the majority of existing methods rely on passive stimuli such as visual and audio, with limited exploration of interactive or active forms of emotion elicitation in VR~\cite{bayro2025systematic}. As a natural extension of VR, interactive VR environments hold promise for enhancing emotional experiences by allowing users to actively engage with the virtual world.

While prior work using virtual environments supports scene-level interaction (\textit{e.g.}, teleportation), the specific role of object-level interaction in modulating emotional responses remains underexplored~\cite{chirico2018designing, somarathna2022virtual, jiang2024immersive}. In particular, research on emotion in VR has focused on measuring responses and designing affective visualizations, yet few studies have systematically varied the interaction to modulate and potentially enhance its effect on emotional engagement. However, previous studies suggest that object-level interaction holds significant potential. For example, object-level interaction enables VR users to reach out and touch objects, making them much more immersed in the game world than traditional screens~\cite{kessing2009services,jones2017disrupting}. The opportunity to touch an object increases the feeling of perceived ownership of that object, and the valuation of the object is also increased when the touch experience provides either neutral or positive sensory feedback~\cite{peck2009effect}. Novel designs for object interaction can significantly enhance perceived fun and user satisfaction~\cite{yu2024object}. Moreover, objects in the virtual world can exert variable impact or control on users' decisions, and thus object-level interaction has an impact on dominance by suggesting or preventing users from accomplishing actions, as does a lack of interaction with the environment~\cite{dozio2022design}.
Despite these insights, it remains unclear how these mechanisms influence emotional experience in VR. Investigating this relationship can clarify the affective role of object-level interaction and guide the design of more emotionally engaging VR environments, which could be applied in mental health research and practice~\cite{spytska2024use,bell2020virtual}.
In this work, we focus on exploring the fundamental question: 
\textit{Does enabling object-level interaction in virtual environments enhance the elicitation of emotional experiences?} 

To investigate this, we extend the VR emotion-elicitation dataset by~\citet{jiang2024immersive} in two ways. First, we integrate affectively-designed interactive elements tailored to each scene by \citet{jiang2024immersive}. For our experimental comparison, each scene is implemented in two versions: an interactive version featuring user-triggered actions (\textit{e.g.}, petting puppies, throwing stones, using a shield) and a non-interactive version with identical audiovisual content.
Furthermore, we extend the existing dataset~\cite{jiang2024immersive} by adding an additional scene. The authors reported that the dataset lacked a scene in the High-Arousal High-Valence (HAHV) quadrant of Russell’s circumplex model~\cite{russell1980circumplex}, which limited its coverage of the full affective space. To address this limitation, we developed a new scene, \textit{Surrounded by Elephants}, based on a validated 360$\degree$ video~\cite{li2017public}. This addition ensures that all four quadrants are now represented, enabling more balanced emotion elicitation. We have implemented both interactive and non-interactive versions of the scene to investigate the effect of object-level interaction on emotion elicitation.

To evaluate the emotional impact of interaction, we adopt a multimodal approach that combines subjective self-reports with objective physiological measures. While the Self-Assessment Manikin (SAM) questionnaire provides valuable subjective ratings of valence and arousal, it has inherent limitations~\cite{bradley1994measuring}. It is a discrete, post-hoc measure that is susceptible to cognitive biases and recall inaccuracies, and it may fail to capture the full intensity of transient, high-arousal states. Physiological sensing, in contrast, provides an objective, continuous, and unconscious measure of affective responses, particularly for arousal-related responses~\cite{sarsenbayeva2019measuring}. This is especially critical for our study, as the visceral, immediate impact of interactive elements -- like the surprise of an elephant's reaction or the tension of raising a shield -- may manifest in the autonomic nervous system before they are consciously processed and reported. By integrating physiological data, we aim to capture these nuanced, real-time reactions, providing a more complete and robust understanding of how interaction modulates emotional experience. This work aims to explore the following research questions:
\begin{itemize}
\item[\textbf{RQ1}] Does the added scene \textit{Surrounded by Elephants} reliably and effectively elicit High Arousal and High Valence emotions?
\item[\textbf{RQ2}] How does object-level interaction influence subjective and physiological measures of emotional response compared to a non-interactive baseline?
\item[\textbf{RQ3}] What is the relationship between subjective self-reports and physiological arousal in response to affective interactions in VR?
\end{itemize}

In summary, our work makes three key contributions to the fields of HCI and affective computing:
\begin{itemize}
\item We provide an extended VR dataset\footnote{\url{https://github.com/ZHEYUANK/VR-Dataset-Emotions-Interaction.git}} with a new scene that fills a critical gap in eliciting high-arousal, high-valence emotions, enabling richer and more balanced affective research.
\item We provide one of the first systematic investigations into the 
effects of object-level interaction on emotional response, using a controlled experimental design with interactive and non-interactive versions of each scene.
\item We provide an empirical analysis combining self-report and physiological data, offering nuanced insights into affective processes and 
propose practical guidance for designing emotionally resonant interactive VR experiences.
\end{itemize}

\section{Related Work}

In this section, we provide an overview of key areas relevant to our work: methods for measuring emotions, approaches to emotion elicitation in VR, and techniques for affective interaction.

\subsection{Measuring Emotions}

The widely used Circumplex Model of Affect~\cite{russell1980circumplex} conceptualizes emotions along two continuous dimensions: valence and arousal. Ekman’s model~\cite{ekman1992argument} classifies emotions into discrete basic categories; however, its universality remains debated due to, e.g., cultural variations~\cite{betella2016affective}. Building on these theoretical frameworks, emotions in HCI research are often measured subjectively using psychometric self-report scales. Common approaches include categorical emotion scales~\cite{csikszentmihalyi1987validity}, the Positive and Negative Affect Schedule (PANAS)~\cite{watson1988development}, the Self-Assessment Manikin (SAM)~\cite{bradley1994measuring}, the Pleasure-Arousal-Dominance (PAD) scale~\cite{mehrabian1996pleasure}, and the Affect Slider~\cite{betella2016affective}. PANAS combines valence and arousal into positive and negative affect scores but may misinterpret pleasure-driven positive affect, especially in high-arousal negative scenarios such as anger, leading to misleadingly high positive scores~\cite{crawford2004positive, pollak2011pam, watson1999two}. SAM introduces a pictorial assessment with pleasure (valence), arousal, and dominance dimensions, providing an intuitive and efficient tool for self-reporting emotions~\cite{bradley1994measuring}. Recent studies have showed its robustness and reliability in VR environment validations~\cite{xie2020applying}. Given its multidimensional nature and image-based icons, SAM is particularly suitable for immersive VR contexts, and we adopt it in our study to capture valence, arousal, and dominance more comprehensively.

Objective emotion measures mostly rely on wearables and physiological sensors, as these provide convenient ways to estimate emotions by measuring multiple signals linked to the central and autonomic nervous systems~\cite{yang2021behavioral}, including electroencephalography (EEG)~\cite{suhaimi2020eeg}, heart rate (HR) and heart rate variability (HRV)~\cite{wascher2021heart}, electrodermal activity (EDA)~\cite{barathi2020affect}, facial behaviors~\cite{sarsenbayeva2020does, yang2021benchmarking, tag2022emotion}, and eye blinks~\cite{zhang2023blink}. However, these signals can be difficult to interpret, as similar patterns may reflect different emotional states and are often influenced by noise and user movement, especially in immersive VR settings~\cite{babaei2021critique, tauscher2019immersive}. Despite offering objective insights, physiological signals do not fully capture the subjective experience of emotion and require careful preprocessing and context-specific interpretation~\cite{potts2024sweating}. To improve reliability, combining multiple physiological signals can enhance emotion recognition accuracy and allow for richer measurement of affective states~\cite{gupta2024caevr}. In addition, VR provides a highly controllable and repeatable setting where such multimodal measurements can be precisely integrated and validated~\cite{li2025encumbrance,li2025wice}, supporting more comprehensive assessments of emotional responses~\cite{li2022neurophysiological, xie2024electroencephalography, li2024multimodal, potts2025retrosketch}.

\subsection{Eliciting Emotions in VR}

Emotion elicitation in VR has traditionally relied on brief and isolated virtual environments designed to induce specific affective states, typically through brief exposures lasting only a few minutes~\cite{estupinan2014can, jicol2021effects, marin2020emotion}. These environments often incorporate visual stimuli such as images from the International Affective Picture System (IAPS)~\cite{lang1995emotion} or 360$\degree$  videos~\cite{schone2023library}, alongside auditory cues like emotionally expressive music~\cite{siedlecka2019experimental}. Beyond perceptual stimuli, recent studies have adapted autobiographical recall into VR, using immersive cue-based paradigms to trigger emotional memories and associated physiological responses~\cite{gupta2022total}.

While effective for controlled induction, these approaches often lack interaction and dynamic affective progression. Recent studies have explored more immersive experiences, particularly VR games, which unfold over extended periods and evoke evolving emotional trajectories~\cite{granato2020empirical, ishaque2020physiological}. These experiences span a range of valence and arousal levels~\cite{bender2021fright, meuleman2018induction}, and can be shaped by design factors such as avatar expressions~\cite{jun2018full}, ambient lighting and color~\cite{bartram2017affective}, and music~\cite{kern2020influence}. However, the complexity and duration of VR games may introduce variability in emotional responses and reduce their suitability for tightly controlled experimental studies~\cite{lavoie2021virtual, lemmens2022fear}.

Beyond basic emotional triggers, recent work has begun designing immersive VR environments to elicit more complex states such as awe~\cite{chirico2018designing} and has provided evidence that emotionally responsive experiences can be achieved through deliberate virtual environment design~\cite{somarathna2022virtual}. As a highly immersive, interactive, and customizable medium, VR offers a promising platform for studying how specific design factors shape emotional experience~\cite{jiang2024immersive}. 

\subsection{Affective Interaction}
Affective interaction refers to the ways in which emotionally meaningful exchanges occur between a user and a designed system~\cite{lottridge2011affective}. While many VR emotion studies rely on passive exposure, recent work has begun to explore how interactive elements shape emotional experience. Studies have shown that features such as haptic feedback~\cite{elor2021understanding}, auditory signals~\cite{halvey2012augmenting,li2025noise}, and interaction with virtual objects~\cite{Liang_2025}, virtual agents~\cite{jacucci2024haptics}, virtual pets~\cite{oxley2022systematic} can modulate affective responses. For example, throwing paper planes in VR can help users symbolically release negative emotions~\cite{wagener2024moodshaper}. Such interactive designs have been shown to enhance emotional engagement by enabling natural and goal-directed actions in immersive VR settings~\cite{somarathna2022virtual,li2025trends}. Furthermore, researchers demonstrated that presence and emotion reinforce each other during interaction~\cite{riva2007affective}, further supporting the role of interaction in affective experience. However, most existing studies are designed for specific applications and lack reusable resources to systematically study how interaction design in VR contributes to emotional responses. Our work examines how affective interaction influence user engagement and emotional connection in VR-based emotion elicitation.

\section{Scene Modeling}

\begin{figure*}[t]
  \centering
  \includegraphics[width=\linewidth]{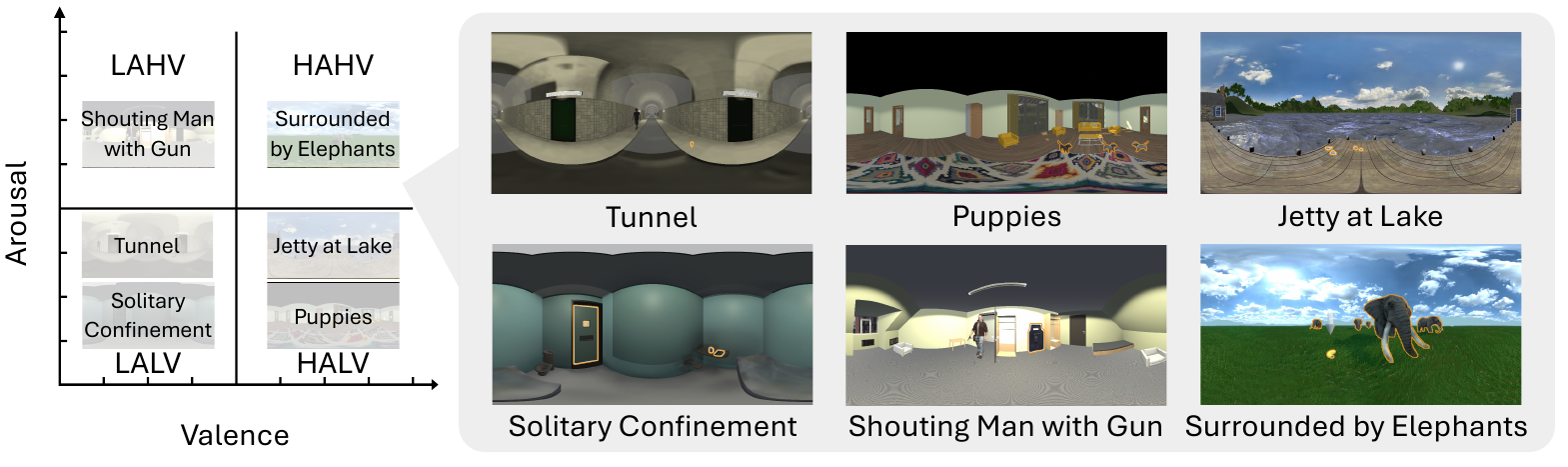}
  \caption{Overview of the six VR scenes. Left: placement of the scenes within Russell’s valence–arousal circumplex model, covering all four quadrants (LAHV, HAHV, LALV, HALV). Right: panoramic views of the scenes with interactive objects highlighted in yellow.}
  \Description{On the left, a valence–arousal plot shows the horizontal axis as valence (increasing to the right) and the vertical axis as arousal (increasing upward), divided into four labeled quadrants (LAHV, HAHV, LALV, HALV). The six scene names are placed in these quadrants, with Shouting Man with Gun in low valence–high arousal, Surrounded by Elephants in high valence–high arousal, Tunnel and Solitary Confinement in low valence–low arousal, and Jetty at Lake and Puppies in high valence–low arousal. On the right, six labeled panoramic scene images (Tunnel, Puppies, Jetty at Lake, Solitary Confinement, Shouting Man with Gun, Surrounded by Elephants) show example viewpoints; interactive objects within each scene are highlighted in yellow.}
  \label{fig:scenes}
\end{figure*}

To understand how interaction affects emotion elicitation in VR, we extend the validated dataset of \citet{jiang2024immersive} by incorporating interactive elements into each scene (details below). Each scene was designed to target a specific quadrant of Russell’s circumplex model of affect~\cite{russell1980circumplex} as shown in \autoref{fig:scenes}: \textit{Jetty at Lake} and \textit{Puppies} fall into the Low Arousal High Valence (LAHV) quadrant, \textit{Tunnel} and \textit{Solitary Confinement} into the Low Arousal Low Valence (LALV) quadrant, and \textit{Shouting Man with Gun} into the High Arousal Low Valence (HALV) quadrant. 
Furthermore, the original dataset did not include a scene targeting the High Arousal High Valence (HAHV) quadrant, which the authors acknowledged as a limitation due to ethical considerations and risks of cybersickness~\cite{jiang2024immersive}. To address these concerns, we ensured that the new HAHV scene was ethically appropriate, designed to elicit strong but positive emotions such as excitement and awe while avoiding distressing or fear content~\cite{linares2024interactive}.

To address this gap, we followed \citet{jiang2024immersive}'s approach and selected a HAHV video clip --- \textit{Surrounded by Elephants}\footnote{Link to the Surrounded by Elephants original video \url{https://www.youtube.com/watch?v=mlOiXMvMaZo}} from \citet{li2017public} consisting of 73 video clips. 
Following the setup of the original video, the user spawns in an open grassland under a bright sky, seeing a herd of seven elephants slowly approaching together. The scene offers a wide, unobstructed view, only limited by hills in the far distance. This video was selected to minimize the risk of cybersickness, as prior studies have shown that VR content with strong and dynamic visual motion (\textit{e.g.}, roller coaster simulations) often provokes cybersickness~\cite{caserman2021cybersickness}, whereas gently paced, naturalistic scenes are expected to pose a lower risk of cybersickness. This consideration is a critical factor in mitigating potential discomfort and ensuring accessibility for a wide range of participants~\cite{jiang2024immersive}.
We then recreated this scene as an immersive VR environment by converting the 360$\degree$ video into a 3D model using Blender and Unity. 

\paragraph{Interactive Scene Design} 

To examine how interaction influences emotional responses, we designed the interactive elements to be intuitive and emotionally meaningful. 
Here, we define interaction as direct manipulation with objects within the scene (\textit{e.g.}, picking up, throwing, touching, or using them).
Each scene was implemented in two versions: an interactive version with user-triggered actions and a non-interactive version with identical audiovisual content, but no object-level interactions. This setup allows controlled comparisons of how interactive elements contribute to emotion elicitation. 

In the interactive scenes, players can teleport and interact with game objects using controllers. All interactive elements in the scenes are highlighted with a yellow outline when the player’s distance to the element is less than 2.5 meters, providing clear visibility and intuitive feedback about potential interactions in a natural reach or teleportation range~\cite{laviola20173d, jerald2015vr}. Each scene also includes corresponding sound effects as detailed in \autoref{tab:scenes}. As illustrated in \autoref{fig:scenes}, the six modeled scenes highlight the interactive objects in yellow. 
The details of the interaction in six scenes are shown in \autoref{fig:objects} in \autoref{app:objects} and described below. 

\paragraph{\textbf{Tunnel}}  
The player finds themselves in a long tunnel lit by dim lights, with pedestrians occasionally passing by. A flashlight lies on the tunnel floor. The player can pick it up. The action triggers light vibrotactile feedback in the controller. Once grabbed, the flashlight turns on and can be used to illuminate various areas of the tunnel.

\paragraph{\textbf{Puppies}}
The player spawns in a spacious and furnished room with three puppies moving around. A tennis ball is placed on a table. The player can pet the puppies to trigger soft vibrotactile feedback, causing them to turn toward the player and sit down. The player can also pick up and throw the tennis ball; the puppies will chase the ball and bring it back. If the player does not interact with the puppies for over 15 seconds, the puppies will sit still on the ground.

\paragraph{\textbf{Jetty at Lake}}  
The player spawns on a jetty in front of a stone house by a calm lake, with hills covered by trees and grass. At the end of the jetty, two stones and two paper planes are placed. The player can grab the stones, triggering short vibrotactile pulse feedback, and throw them into the lake, which produces splash sounds and visible ripple effects on the water surface. Paper planes can also be picked up, triggering light vibrotactile feedback, and thrown, with a visible trajectory rendered at the tail during flight.

\paragraph{\textbf{Solitary Confinement}} 
The player spawns in a confined, gloomy cell with a flashing light, a toilet seat, and a single bed. A book and a metal cup are placed on the table. The player can knock on the iron door, triggering strong vibrotactile feedback and loud knocking sounds. The cup and book can be picked up and thrown at the door.

\paragraph{\textbf{Shouting Man with Gun}} 
The player spawns inside a furnished attic, where, after a short delay, a man breaks in shouting and aiming a pistol at them. A metal riot shield is placed against the wall, featuring a small bulletproof glass window. The player can grab the shield for protection. The shield can be used to block the view, and the player can observe the man through the window.

\paragraph{\textbf{Surrounded by Elephants}}  
The player is placed in a green grassland with distant hills and a cloudy sky, where a herd of elephants slowly approaches. A banana floats above the grass. The player can grab the banana, trigger light vibrotactile feedback, and throw it toward the elephants. This triggers the elephant to pick it up with its trunk and eat it. When the player, touches the elephant, it triggers deep vibrotactile feedback in the controller; the elephant takes two steps back, raises its trunk, and produces a vocal sound.


We developed our scenes using Unity 6000.0.46f1 Long-Term Support (LTS) version to ensure compatibility with the Meta Quest runtime environment and the SDKs employed in this study. All 3D models were either imported from open-source repositories licensed under CC BY-SA, obtained from the Unity Asset Store under its standard license, or created using \href{https://www.blender.org/}{Blender}.

\begin{table*}[t]
  \caption{Details of the modeled interactive scenes for emotion elicitation. \textit{V -- Valence, A -- Arousal}.}
  \label{tab:scenes}
  \begin{tabularx}{\linewidth}{p{1.8cm}Xp{2.3cm}p{2.8cm}p{2.0cm}l}
    \toprule
    \textbf{Scene Name} & \textbf{Visual Description} & \textbf{Sound \newline Effect} & \textbf{Interactive \newline Elements} & \textbf{Targeted \newline Emotion} & \textbf{Ref.} \\
    \midrule
    Tunnel & A long tunnel lit by dim lights, with pedestrians occasionally passing by. & Footsteps & Picking up a flashlight. & V: Mid-to-low \newline A: Mid-to-low & \cite{schone2023library} \\
    \midrule
    Puppies & A spacious and furnished room with several puppies around. & (Quiet) & Petting puppies and throwing ball. & V: High \newline A: Low & \cite{schone2023library, li2017public} \\
    \midrule
    Jetty at Lake & A jetty in front of a stone house by a lake, with hills covered by trees and grass. & Water flow & Throwing stones and paper airplanes. & V: High \newline A: Low & \cite{schone2023library} \\
    \midrule
    Solitary \newline Confinement & A gloomy cell with a flashing light, a toilet set, and a single bed. & Water \newline dropping & Knocking door; grabbing a book and a cup. & V: Low \newline A: Low & \cite{li2017public} \\
    \midrule
    Shouting Man with Gun & A furnished attic. A furnished attic where a man breaks in while shouting and aims a pistol at the player after a certain time. & Man \newline shouting & Using shield to block and observe. & V: Mid-to-low \newline A: High & \cite{schone2023library} \\
    \midrule
    Surrounded by Elephants & A green grassland with distant hills and a cloudy sky, where elephants walk toward. & Wind and elephant trumpeting & Feeding and touching elephants & V: High \newline A: High & \cite{li2017public} \\
    \bottomrule
  \end{tabularx}
\end{table*}

\section{Study 1: Validating the effectiveness of the \textit{Surrounded by Elephants} Scene}
First, we carried out a user study following a within-subjects design to evaluate if the VR scene \textit{Surrounded by Elephants} elicited HAHV emotions as effectively as its original 360$\degree$ video format (RQ1). The study was approved by our university's research ethics committee. 

\subsection{Apparatus}
The study took place in a university lab with an unobstructed room-scale tracking area of more than $3\times3$ metres. We used a Meta Quest Pro VR headset for the study. We rendered the scene in Unity and ran it on a desktop PC (Intel Core Ultra 9 285K, 32GB DDR5 RAM, Nvidia RTX 4080 SUPER) connected via a Quest Link cable. 

\subsection{Procedure}
Upon arrival in our lab, we assigned each participant a unique anonymous identifier for data management. Moreover, we screened participants for potential health risks related to VR use (\textit{e.g.}, epilepsy, mobility or visual impairments, adverse emotional responses). Eligible participants received an overview of the study, provided informed consent, and completed a questionnaire on demographics and prior VR experience. Participants then completed a tutorial scene to familiarize themselves with the VR system, during which a baseline SAM questionnaire~\cite{bradley1994measuring} was administered. Participants subsequently experienced one 360$\degree$ video and one VR scene in a balanced random order, each lasting at least 30 seconds with the option to continue exploring before completing the SAM scale (see \autoref{fig:sam}). Between scenes, participants transitioned back to the lab environment via the headset's mixed reality feature for a 60-second break. The study lasted approximately 10 minutes per participant.

\subsection{Participants}
We recruited 24 volunteers (12 female, 12 male) aged between 19 and 38 years (M = 25.63, SD = 3.70) to participate in our study. All participants provided informed consent prior to the study. They did not receive monetary compensation, but were offered snacks. Our participants reported different levels of VR experience: 8 used VR daily, 7 weekly, 5 monthly, and 4 never.

\subsection{Results}
We first evaluated how effectively the \textit{Surrounded by Elephants} VR scene elicited emotion. We collected a total of 72 SAM measurements (24 participants $\times$ 2 \textsc{Conditions} $\times$ \textit{Surrounded by Elephants}). The two \textsc{conditions} were the original 360$^{\circ}$ video~\cite{li2017public} and the respective modeled VR scene. 

After checking for normality, we applied ART ANOVAs~\cite{wobbrock2011aligned} to analyze the non-normally distributed data (Shapiro-Wilk test, $p < 0.05$). For Valence, the main effect of \textsc{Condition} was significant, $F(1, 23) = 15.597$, $p < 0.001$, $\eta^2 = 0.440$, indicating that the \textit{Surrounded by Elephants} VR scene elicited significantly higher ratings than the 360$^{\circ}$ Video. For Arousal, the main effect of \textsc{Condition} was not significant, $F(1, 23) = 0.055$, $p = 0.817$, $\eta^2 = 0.004$, indicating no difference between the VR Scene and the 360$^{\circ}$ Video. For Dominance, the main effect was significant, $F(1, 23) = 9.015$, $p < 0.01$, $\eta^2 = 0.368$, with higher ratings for the VR scene compared to the 360$^{\circ}$ video. Overall, the VR scene elicited higher valence and dominance but similar arousal compared to its 360$^{\circ}$ video counterpart, consistent with previous findings. Our results show that our \textit{Surrounded By Elephants} VR Scene can elicit high-valence, high-arousal emotion. \autoref{tab:elephant_sam} in \autoref{app:tables} presents the results, showing that the VR scene elicited emotional responses comparable to the 360$^{\circ}$ video, indicating its effectiveness for high-arousal, high-valence emotion elicitation.

\section{Study 2: Understanding Effects of Interaction on Emotion Elicitation}
Then we conducted a mixed-design user study to investigate the role of affective interaction when eliciting emotions in VR (RQ2 and RQ3). Affective interaction design (Interactive vs. Non-Interactive) served as a between-subject factor, while the six VR scenes were used as a within-subject factor. We applied a balanced Latin square to counterbalance the scene order across participants. Physiological and self-report measures were collected throughout to examine how interactive elements influence emotional responses and how these responses are reflected in physiological signals. This user study received ethics approval from our university.

\subsection{Apparatus}
\begin{figure}[t]
  \centering
  \includegraphics[width=\linewidth]{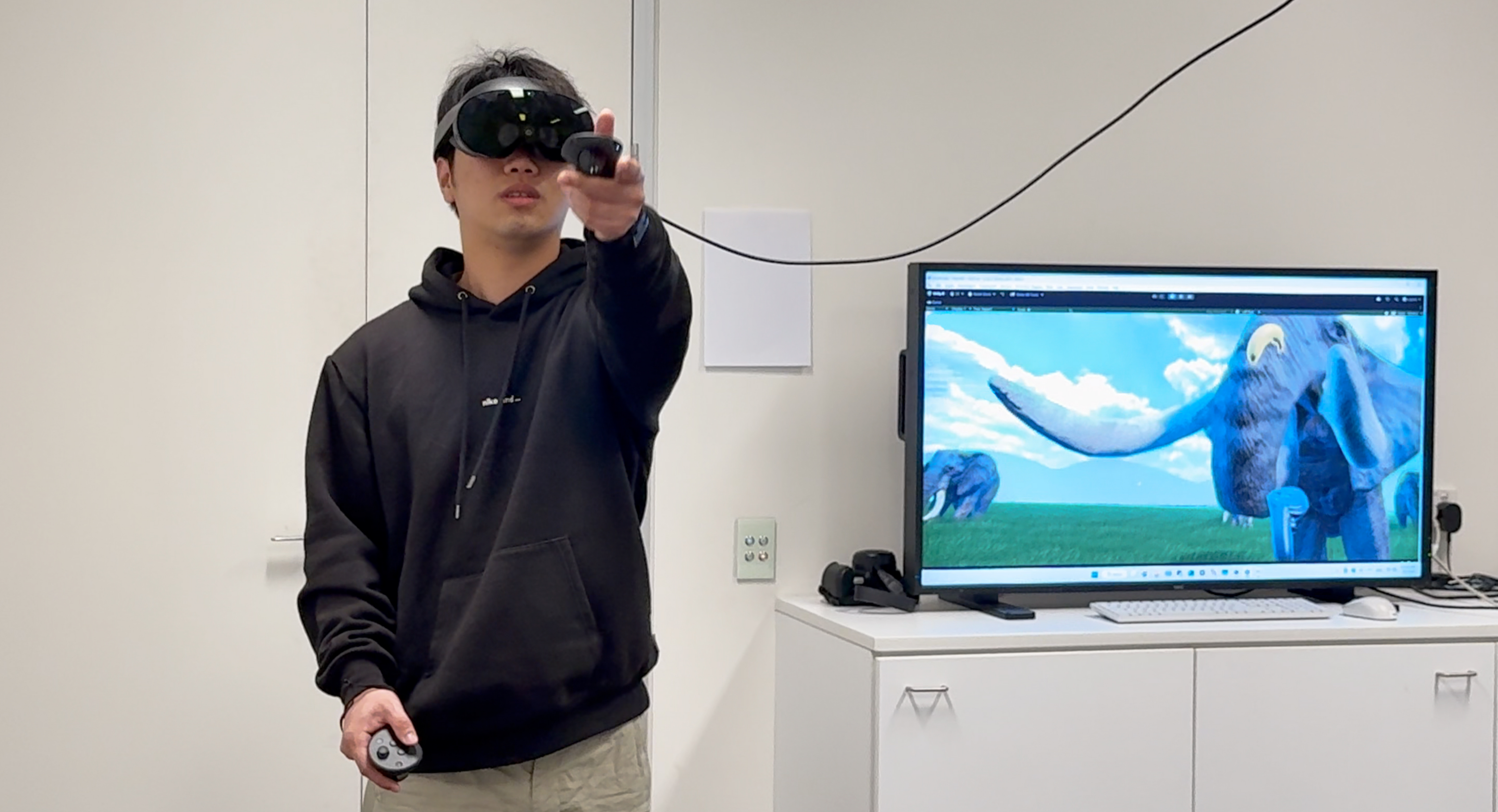}
  \Description{A participant stands in an indoor lab space wearing a VR headset and holding two handheld controllers, with one arm extended forward as if pointing or selecting an object. A tethered cable runs from the headset toward the ceiling. To the participant’s right, a large monitor on a cabinet displays the VR scene being experienced (showing an outdoor environment with an elephant), providing an external view of the participant’s in-headset content.}
  \caption{An example of the experimental setup for participant experiencing VR scenes.}
  \label{fig:Participant}
\end{figure}

We used a Meta Quest Pro VR headset and two Meta Quest Touch Pro controllers for all VR scenes. The study was conducted in a university lab space with an unobstructed room-scale tracking area of over $3\times3$ meters to support full-body interaction. All scenes were rendered in Unity and ran on a desktop PC (Intel Core Ultra 9 285K, 32GB DDR5 RAM, Nvidia RTX 4080 SUPER) via a Quest Link cable connection. The experimental setup is shown in \autoref{fig:Participant}.

\subsection{Measurements}
During the study, we used the SAM questionnaire to assess participants' emotional states. We also collected their physiological data using the \href{https://www.empatica.com/en-int/embraceplus/}{Empatica EmbracePlus} wristband during the study. We recorded electrodermal activity (EDA) and blood volume pulse (BVP) signals via the \href{https://www.empatica.com/embraceplus-compatibility/}{Empatica Care Lab} mobile application. EDA was sampled at 4 Hz, and BVP at 64 Hz. In addition, we conducted semi-structured interviews in which participants identified the scenes they found most and least emotionally impactful and provided explanations for their choices.

\subsection{Participants}
We recruited 84 participants (42 female, 42 male) aged between 19 to 35 years (M = 24.05, SD = 3.03). All participants provided informed consent prior to the study and were compensated with a gift card ($\approxeq15USD$) for their participation. Each participant was assigned a unique anonymous identifier for data management. Participants came with various VR usage experiences, with 14 reporting daily use, 21 weekly, 25 monthly, and 24 never. \autoref{tab:demographics} in \autoref{app:tables} presents the demographic details for the user study.

\begin{figure}[t]
  \centering
  \includegraphics[width=\linewidth]{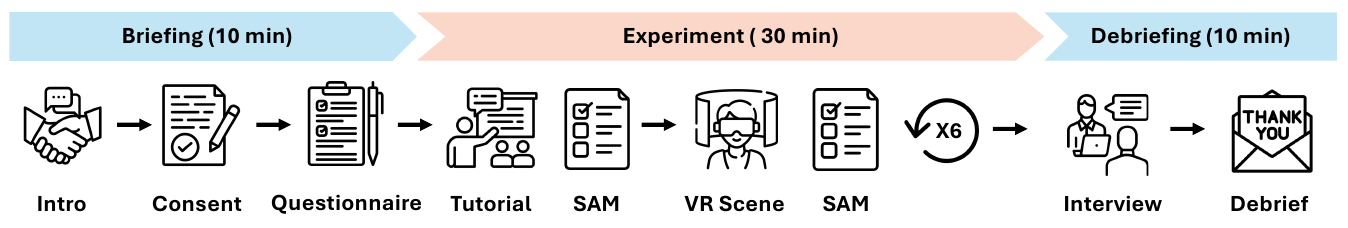}
  \Description{The diagram shows three phases arranged left to right: Briefing (10 min), Experiment (30 min), and Debriefing (10 min). Under Briefing, icons indicate an introduction, consent, questionnaire, and tutorial. Under Experiment, participants complete a Self Assessment Manikin (SAM) rating, experience a VR scene, and complete another SAM rating; this scene plus rating sequence is repeated six times. Under Debriefing, the steps are an interview followed by a final debrief/thank you. Steps conducted outside VR are colored blue, while steps conducted in VR are colored orange.}
  \caption{Study Protocol: Some study elements were conducted outside the VR environment (blue), while the main study components occurred within the VR environment to avoid breaking participants' immersion (orange).}
  \label{fig:protocol}
\end{figure}

\subsection{Procedure}
The study protocol is illustrated in \autoref{fig:protocol}. Upon arrival in our lab, participants were first screened for potential health risks related to VR use, including epilepsy, mobility impairments, severe visual impairments, and a history of adverse emotional responses such as anxiety disorders or PTSD through a questionnaire. We then provided eligible participants with an overview of the study’s purpose. After confirming their understanding, they provided written consent and completed a 
questionnaire that collected demographic information, including gender, age and prior VR experience. We then fitted the participants with an EmbracePlus wristband to record physiological data, which they wore for the entire duration of the study, approximately 50-60 minutes per participant.
After that, we asked participants to complete a tutorial, designed to introduce them to the VR system and ensure their familiarity with the headset and control methods, including teleportation for navigation and interaction with the designated virtual objects. 

The tutorial scene was implemented in two versions to familiarize participants with the VR system. Both versions were set in an open, unobstructed virtual space with a text panel in front of the participant explaining the study procedure and controls. In the non-interactive version, participants practiced teleportation by moving to an exit point and then completed the SAM questionnaire~\cite{bradley1994measuring}. In the interactive version, two blue cubes were placed in front of the participant, and participants practiced picking them up and throwing them before teleporting to the exit point to complete the SAM questionnaire. Within this scene, we also collected a baseline measurement of their emotional state using the SAM questionnaire. 

Once comfortable with the setup, participants proceeded to experience all six VR scenes in a counterbalanced order using a Latin square design. Each scene lasted for at least 30 seconds, after which an exit point appeared. Participants could, however, choose to remain in the scene for further exploration or leave the scene by teleporting to the exit point. Upon leaving, they were teleported to the initial position of the scene to complete the SAM questionnaire. Afterwards, they exited the scene and returned to the mixed reality environment of the lab for a 60-second break to reduce any potential emotional carryover.

After completing all scenes, participants exited the VR application and took part in a semi-structured interview. This final phase aimed to gather qualitative insights into their emotional responses and overall experiences with each scene.

\section{Evaluation Results}
To capture how object-level interaction shapes emotional experience in VR, we detail the findings of our study in this section and provide a comprehensive evaluation of the interaction of the scenes. Precisely, we first compared SAM results between the interactive and non-interactive versions to examine differences in emotional elicitation. To uncover the behavioral and contextual mechanisms behind these ratings and gain a more comprehensive understanding of the emotion elicitation process, we further analyzed participants’ interactions with the virtual environment, including \textit{time} engaged with each scene, \textit{virtual position} and \textit{orientation}, and emotion-related \textit{descriptions} extracted through topic modeling. In addition, we analyzed physiological data, including BVP and EDA, to complement the assessment of emotional responses.

\subsection{Emotion Elicitation Measurements}
We analyzed 588 SAM measurements (42 participants per group $\times$ 2 \textsc{Affective Interaction Design} $\times$ \textsc{Scene} (6 scenes, and the tutorial as baseline)). The \textsc{Affective Interaction Design} refers to the interactive and non-interactive versions of the VR scenes. As the data deviated from normality (Shapiro–Wilk test, $p < .05$) and the study used a mixed design with repeated measurements across scenes, we used linear mixed-effects models to analyze SAM ratings, focusing on between-group effects. The model specified \textsc{Affective Interaction Design} and \textsc{Scene} as fixed effects with their interaction, included Gender, Age, and XR experience as control variables, and added a random intercept for participants to account for within-subject dependency. The fixed-effects estimates are summarized in \autoref{tab:fixed_effects}.

\begin{figure}[t]
  \centering
  \includegraphics[width=\linewidth]{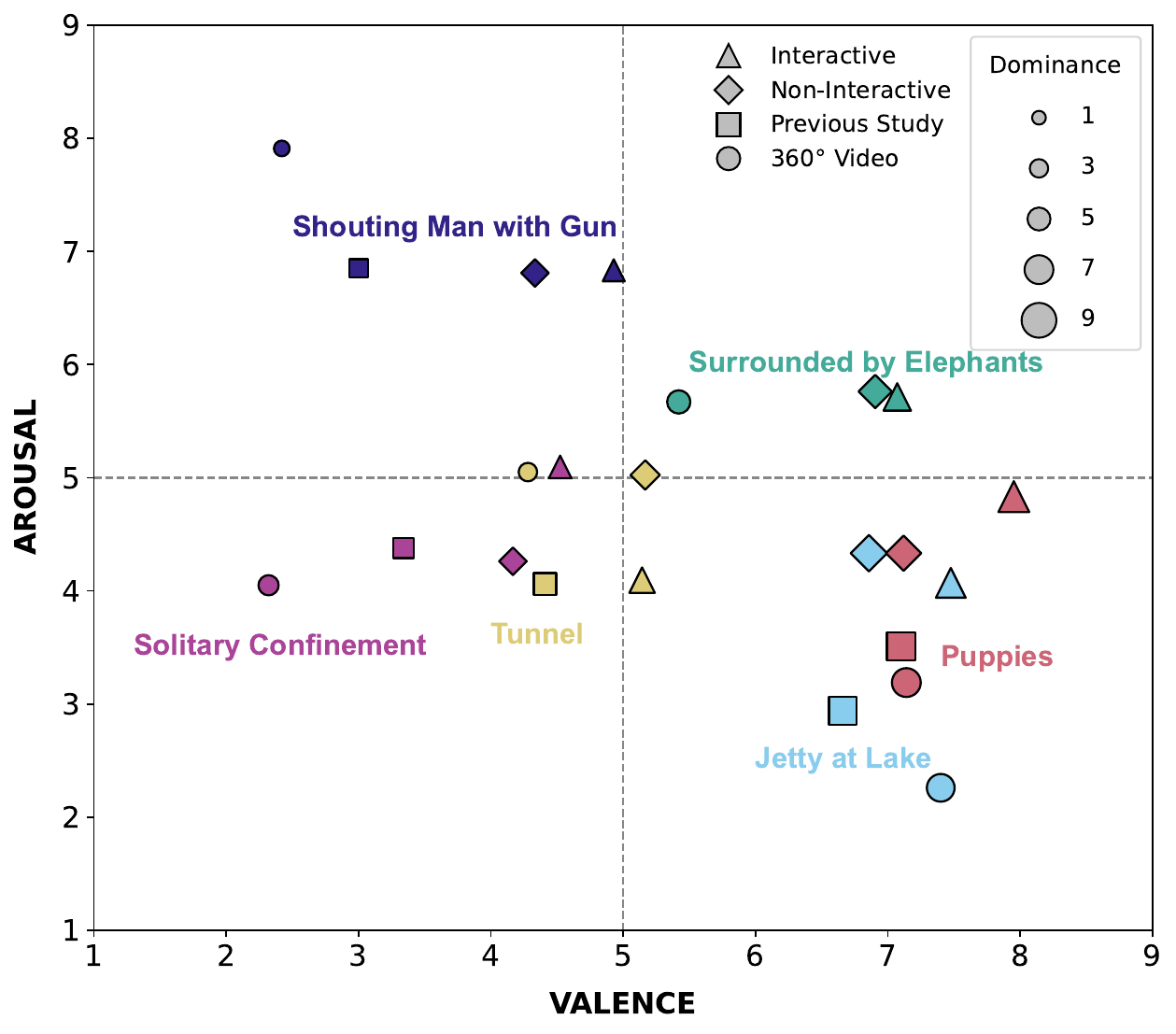}
  \caption{Illustrations of all emotional responses in the valence-arousal space. The x-axis represents valence, the y-axis represents arousal, and marker size represents dominance. Interactive and Non-Interactive results are from this study, Previous Study results from \citet{jiang2024immersive}, and 360$\degree$ results from established datasets~\cite{li2017public, schone2023library}. The \textit{Surrounded by Elephants} scene is not included in the Previous Study.}
  \Description{The x axis is Valence (1–9) and the y axis is Arousal (1–9), with dashed reference lines at valence = 5 and arousal = 5. Each scene is labeled near a cluster of points: Shouting Man with Gun appears in the upper left (low valence, high arousal), Solitary Confinement is left of center at moderate arousal (low valence), Tunnel is near the center around neutral valence and moderate arousal, Surrounded by Elephants is in the upper right (high valence, high arousal), and Jetty at Lake and Puppies are in the lower right (high valence, lower arousal). Marker shape indicates study condition (triangles for interactive, diamonds for non interactive, squares for a previous study, circles for a 360 video), and marker size encodes dominance on a 1–9 scale, with larger symbols indicating higher dominance.}
  \label{fig:Emotional_Space_Comparison} 
\end{figure}

\begin{figure*}[t]
  \centering
  \includegraphics[width=\linewidth]{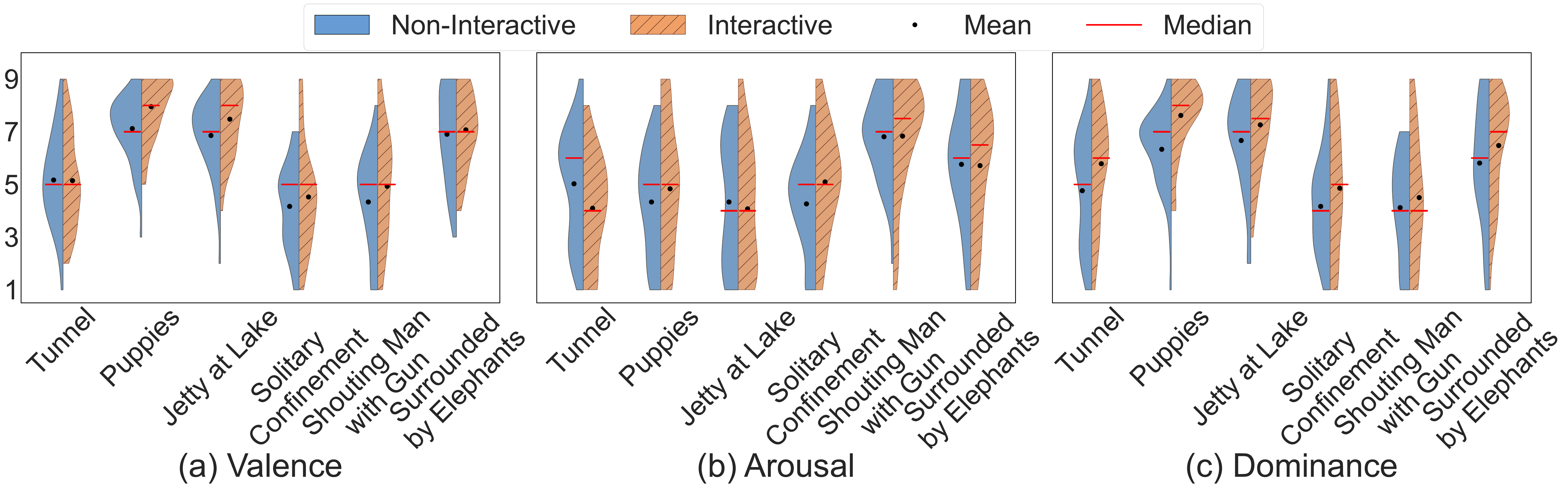}
  \caption{SAM questionnaire results after experiencing each scene for valence, arousal, and dominance.}
  \Description{Panel (a) shows valence, panel (b) shows arousal, and panel (c) shows dominance, each on a 1–9 scale. For each scene on the x axis (Tunnel, Puppies, Jetty at Lake, Solitary Confinement, Shouting Man with Gun, Surrounded by Elephants), two half violins are shown: non interactive in solid blue and interactive in orange with diagonal hatching. Small black markers indicate the mean for each condition, and a short red horizontal line indicates the median. Across panels, Puppies and Jetty at Lake generally appear higher in valence than Solitary Confinement and Shouting Man with Gun, while Shouting Man with Gun tends to show higher arousal than the calmer scenes.}
  \label{fig:SAM-violin}
\end{figure*}

\begin{figure}[t]
  \centering
  
  \begin{subfigure}[t]{\linewidth}
    \centering
    \includegraphics[width=\linewidth]{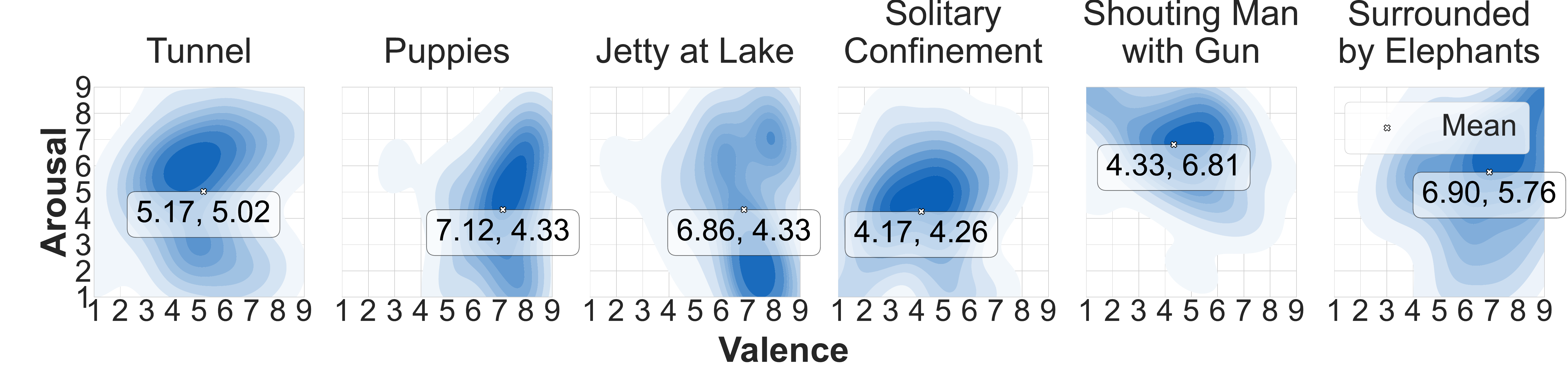}
    \caption{Non-interactive VR scenes}
    \Description{Kernel density estimate plots of valence and arousal values for non-interactive VR scenes.}
    \label{fig:Non-Interactive_Combined_KDE_Plot}
  \end{subfigure}
  
  \begin{subfigure}[t]{\linewidth}
    \centering
    \includegraphics[width=\linewidth]{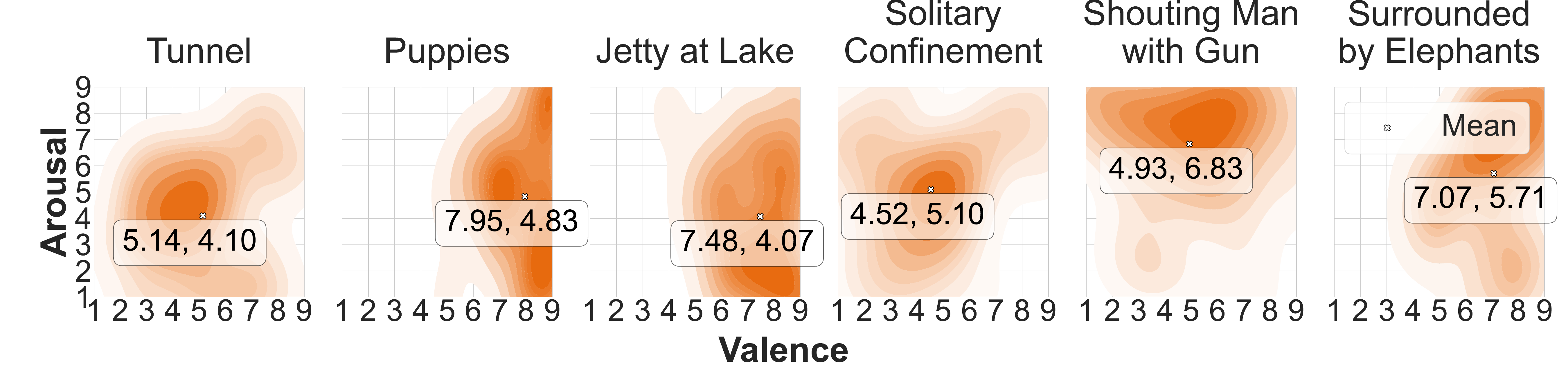}
    \caption{Interactive VR scenes}
    \Description{Kernel density estimate plots of valence and arousal values for interactive VR scenes.}
    \label{fig:Interactive_Combined_KDE_Plot}
  \end{subfigure}
  
  \caption{Kernel density estimate plots of valence and arousal values.}
  \Description{The figure has two rows. Row (a) shows non interactive scenes in blue, and row (b) shows interactive scenes in orange. Within each row, six small square panels correspond to Tunnel, Puppies, Jetty at Lake, Solitary Confinement, Shouting Man with Gun, and Surrounded by Elephants. In each panel, the x axis is Valence (1–9) and the y axis is Arousal (1–9); shaded contours indicate where ratings are most concentrated. Each scene panel includes a small label box reporting the mean valence and mean arousal for that scene. At the far right of each row, an additional panel summarizes the overall mean valence and arousal across all scenes for that condition.}
  \label{fig:KDE_Comparison}
\end{figure}

The emotional responses in the valence-arousal space are illustrated in \autoref{fig:Emotional_Space_Comparison}, \autoref{fig:SAM-violin}, and \autoref{fig:KDE_Comparison}. The results indicate differences in emotional responses based on the presence of interactive elements. For example, in the \textit{Puppies} scene, the interactive version elicited higher valence, arousal, and dominance ratings than the non-interactive version. In the \textit{Shouting Man with Gun} scene, interaction led to higher valence and dominance, while arousal remained similarly high across both versions.

\begin{table*}[t]
    \centering
    \caption{Fixed-effects estimates from linear mixed-effects models for Valence, Arousal, and Dominance. Each outcome was modeled as \textit{Outcome} $\sim$ Affective Interaction Design * Scene + Gender + Age + XR~Experience + (1|Participant), where \textit{Affective Interaction Design} refers to interactive vs.\ non-interactive, and \textit{Scene} includes one tutorial (baseline) and six VR scenes. Values are reported as Estimate (SE) with 95\% CI. }
    \label{tab:fixed_effects}
\begin{tabular}{lcd{0.2}ccd{0.2}ccd{0.2}c}
\toprule
\multirow{2}{*}{\textbf{Parameter}} & 
\multicolumn{3}{c}{\textbf{Valence}} & 
\multicolumn{3}{c}{\textbf{Arousal}} & 
\multicolumn{3}{c}{\textbf{Dominance}} \\
\cmidrule(lr){2-4}\cmidrule(lr){5-7}\cmidrule(l){8-10}
 & \multicolumn{2}{c}{Estimate} & & \multicolumn{2}{c}{Estimate} & & \multicolumn{2}{c}{Estimate} &  \\
 \cmidrule(lr){2-3}\cmidrule(lr){5-6}\cmidrule(l){8-9}
 & M & \multicolumn{1}{c}{SE} & 95\% CI & M &  \multicolumn{1}{c}{SE} & 95\% CI & M &  \multicolumn{1}{c}{SE} & 95\% CI\\
\midrule
Intercept                      & \textbf{6.37\textsuperscript{***}} & 0.92 & [4.54, 8.20] & \textbf{5.80\textsuperscript{***}} & 1.21 & [3.39, 8.20] & \textbf{7.58\textsuperscript{***}} & 1.13 & [5.33, 9.83] \\
Interactive                    & -0.01 & 0.36 & [-0.71, 0.70] & 0.06 & 0.47 & [-0.87, 0.98] & -0.29 & 0.44 & [-1.15, 0.57] \\
Gender (Male)                  & 0.26 & 0.22 & [-0.18, 0.70] & \textbf{-0.85\textsuperscript{**}} & 0.29 & [-1.43, -0.27] & 0.46 & 0.27 & [-0.09, 1.00] \\
Age                            & 0.02 & 0.04 & [-0.05, 0.09] & -0.04 & 0.05 & [-0.13, 0.06] & -0.02 & 0.05 & [-0.11, 0.07] \\
VR Experience                  & -0.07 & 0.11 & [-0.28, 0.14] & 0.26 & 0.14 & [-0.02, 0.53] & -0.24 & 0.13 & [-0.50, 0.01] \\
Tunnel                         & \textbf{-1.76\textsuperscript{***}} & 0.3 & [-2.35, -1.17] & 0.26 & 0.39 & [-0.51, 1.03] & \textbf{-2.21\textsuperscript{***}} & 0.36 & [-2.93, -1.50] \\
Puppies                        & 0.19 & 0.3 & [-0.40, 0.78] & -0.43 & 0.39 & [-1.20, 0.34] & -0.64 & 0.36 & [-1.35, 0.07] \\
Jetty at Lake                  & -0.07 & 0.3 & [-0.66, 0.52] & -0.43 & 0.39 & [-1.20, 0.34] & -0.31 & 0.36 & [-1.02, 0.40] \\
Solitary Confinement           & \textbf{-2.76\textsuperscript{***}} & 0.3 & [-3.35, -2.17] & -0.50 & 0.39 & [-1.27, 0.27] & \textbf{-2.81\textsuperscript{***}} & 0.36 & [-3.52, -2.10] \\
Shouting Man with Gun          & \textbf{-2.60\textsuperscript{***}} & 0.3 & [-3.18, -2.01] & \textbf{2.05\textsuperscript{***}} & 0.39 & [1.28, 2.81] & \textbf{-2.86\textsuperscript{***}} & 0.36 & [-3.57, -2.15] \\
Surrounded by Elephants        & -0.02 & 0.3 & [-0.61, 0.56] & \textbf{1.00\textsuperscript{*}} & 0.39 & [0.23, 1.77] & \textbf{-1.17\textsuperscript{**}} & 0.36& [-1.88, -0.46] \\
Interactive $\times$ Tunnel    & 0.02 & 0.42 & [-0.81, 0.86] & \textbf{-1.14\textsuperscript{*}} & 0.55 & [-2.23, -0.06] & \textbf{1.48\textsuperscript{**}} & 0.51 & [0.47, 2.48] \\
Interactive $\times$ Puppies   & \textbf{0.88\textsuperscript{*}} & 0.42 & [0.05, 1.71] & 0.29 & 0.55 & [-0.80, 1.37] & \textbf{1.74\textsuperscript{***}} & 0.51 & [0.73, 2.74] \\
Interactive $\times$ Jetty at Lake        & 0.67 & 0.42 & [-0.17, 1.50] & -0.48 & 0.55 & [-1.56, 0.61] & \textbf{1.05\textsuperscript{*}} & 0.51 & [0.04, 2.05] \\
Interactive $\times$ Solitary Confinement & 0.40 & 0.42 & [-0.43, 1.24] & 0.62 & 0.55 & [-0.47, 1.70] & \textbf{1.14\textsuperscript{*}} & 0.51 & [0.14, 2.15] \\
Interactive $\times$ Shouting Man with Gun & 0.64 & 0.42 & [-0.19, 1.47] & -0.19 & 0.55 & [-1.28, 0.89] & 0.83 & 0.51 & [-0.17, 1.84] \\
Interactive $\times$ Surrounded by Elephants & 0.21 & 0.42 & [-0.62, 1.05] & -0.26 & 0.55 & [-1.35, 0.82] & \textbf{1.12\textsuperscript{*}} & 0.51 & [0.11, 2.12] \\
\bottomrule
\end{tabular}
\vspace{0.5ex}
\raggedright\footnotesize
\textit{Notes.} Stars denote significance: \textsuperscript{*}\,$p<.05$, 
\textsuperscript{**}\,$p<.01$, 
\textsuperscript{***}\,$p<.001$. 
\end{table*}

\subsubsection{Effects of interaction}
As illustrated in \autoref{tab:fixed_effects}, across all scenes, the main effect of the interaction was not significant: Valence (Estimate = -0.01, SE = 0.36, 95\% CI = [-0.70, 0.71]), Arousal (Estimate = 0.06, SE = 0.47, 95\% CI = [-0.87, 0.98]), and Dominance (Estimate = -0.29, SE = 0.44, 95\% CI = [-1.15, 0.57]). By contrast, the effects of the scene were robust, eliciting the targeted emotions, \textit{e.g.}, reduced Valence and increased Arousal in \textit{Shouting Man with Gun}, or decreased Valence and Dominance in \textit{Solitary Confinement}.

\begin{figure*}[t]
    \centering
    \begin{subfigure}{0.33\linewidth}
        \centering
        \includegraphics[width=\linewidth]{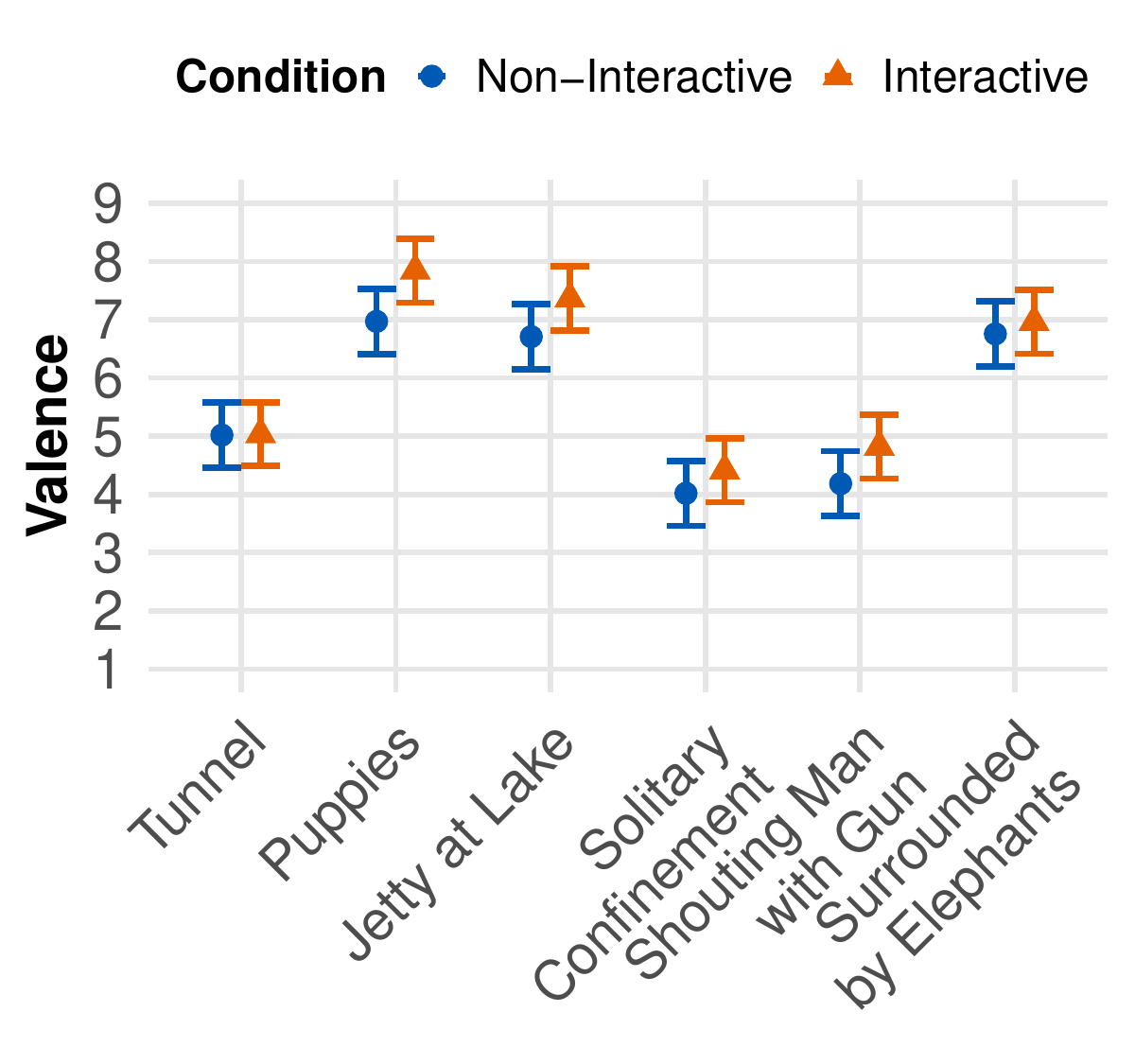}
        \caption{Valence}
        \label{fig:Valence_Lmer}
    \end{subfigure}
    \hfill
    \begin{subfigure}{0.33\linewidth}
        \centering
        \includegraphics[width=\linewidth]{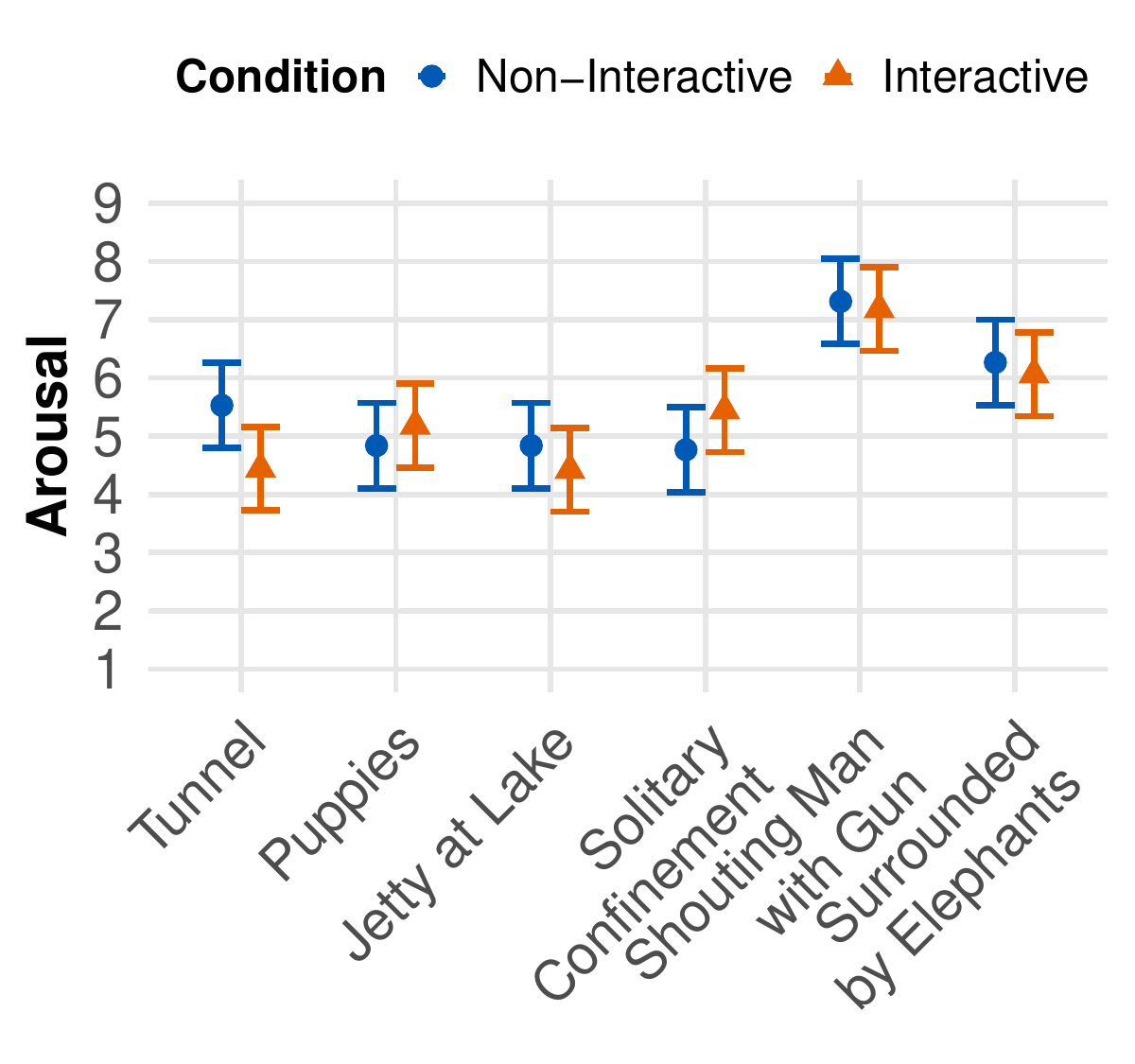}
        \caption{Arousal}
        \label{fig:Arousal_Lmer}
    \end{subfigure}
    \hfill
    \begin{subfigure}{0.33\linewidth}
        \centering
        \includegraphics[width=\linewidth]{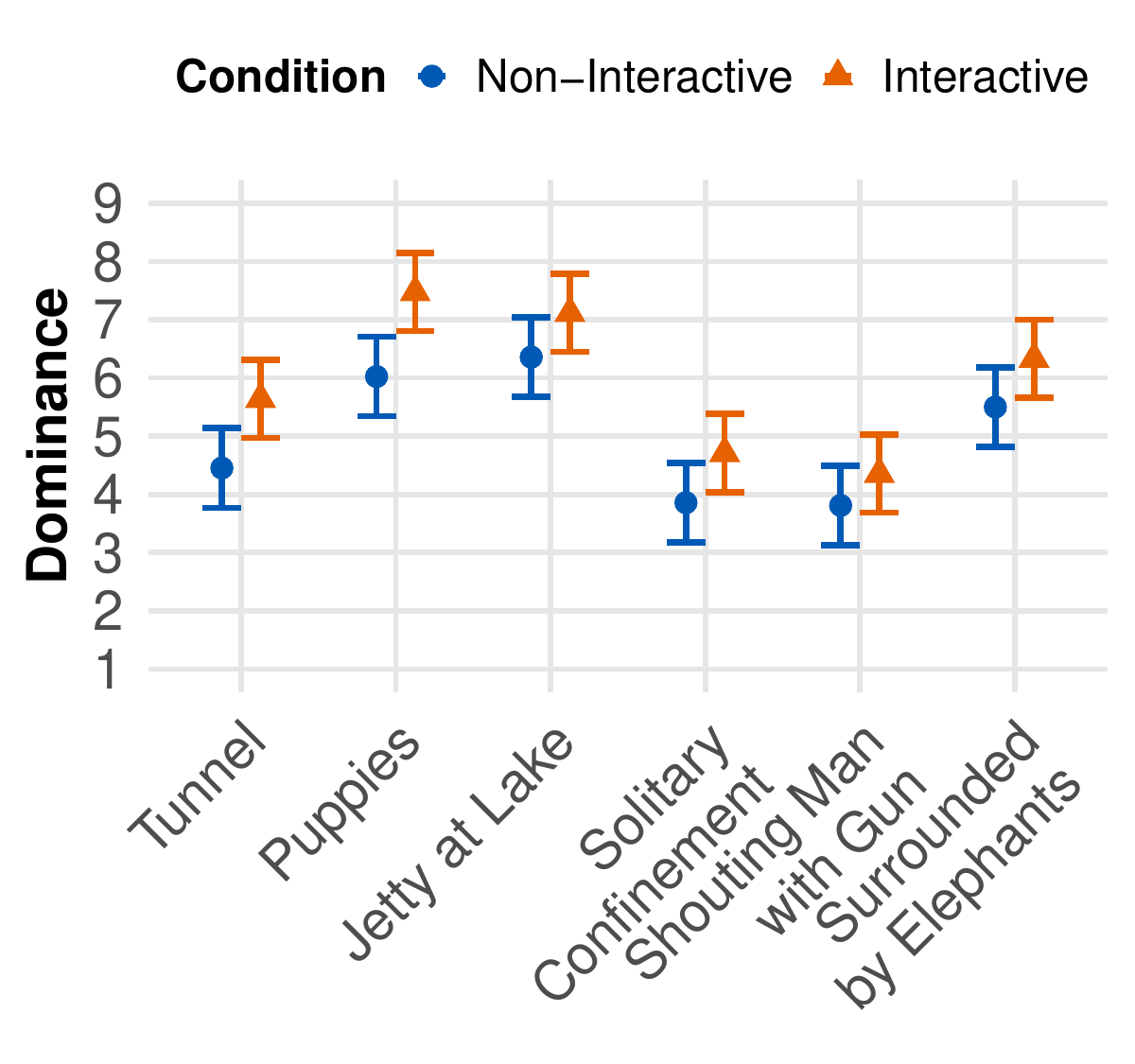}
        \caption{Dominance}
        \label{fig:Dominance_Lmer}
    \end{subfigure}
    \caption{Model-predicted interaction effects from the mixed-effects models for (a) Valence, (b) Arousal, and (c) Dominance across interactive and non-interactive conditions. Each point represents the estimated marginal mean for a given scene, with error bars showing the 95\% confidence interval.}
    \label{fig:Lmer}
    \Description{Panels (a), (b), and (c) correspond to Valence, Arousal, and Dominance, each on a 1–9 scale. The x axis in each panel lists the six scenes (Tunnel, Puppies, Jetty at Lake, Solitary Confinement, Shouting Man with Gun, Surrounded by Elephants). For each scene, two points are plotted: a blue marker for the non interactive condition and an orange marker for the interactive condition. Vertical error bars show the 95\% confidence interval around each estimated mean. Overall, Puppies and Jetty at Lake have higher estimated valence than Solitary Confinement and Shouting Man with Gun, and Shouting Man with Gun has the highest estimated arousal across scenes.}
\end{figure*}

The interaction between \textsc{Affective Interaction Design} and \textsc{Scene} revealed how affective interaction influenced experiences in specific contexts. Valence increased in \textit{Puppies} (Estimate = 0.88, SE = 0.42, 95\% CI = [0.05, 1.71]), with participants rating higher values when they could pet and play the ball with the puppies. Arousal decreased in \textit{Tunnel} (Estimate = -1.14, SE = 0.55, 95\% CI = [-2.23, -0.06]) when participants could use the flashlight. Dominance increased in \textit{Puppies} (Estimate = 1.74, SE = 0.51, 95\% CI = [0.73, 2.74]), \textit{Tunnel} (Estimate = 1.48, SE = 0.51, 95\% CI = [0.47, 2.48]), \textit{Solitary Confinement} (Estimate = 1.14, SE = 0.51, 95\% CI = [0.14, 2.15]), \textit{Jetty at Lake} (Estimate = 1.05, SE = 0.51, 95\% CI = [0.04, 2.05]), and \textit{Surrounded by Elephants} (Estimate = 1.12, SE = 0.51, 95\% CI = [0.11, 2.12]). \autoref{fig:Lmer} illustrates these differences, showing that interaction did not uniformly shift responses, but produced scene-dependent effects, most consistently in measures of Dominance, i.e., the sense of control a user feels over a situation during an emotional experience. Covariates further revealed that male participants reported lower Arousal (Estimate = -0.85, SE = 0.29, 95\% CI = [-1.43, -0.27]) than female participants.

\autoref{fig:Emotional_Space_Comparison} shows scene-specific differences. Interaction generally yielded higher valence and dominance in \textit{Puppies}, \textit{Surrounded by Elephants}, \textit{Jetty at Lake} scenes. In these scenes, participants engaged more, petting the puppies and playing fetch with them, feeding the elephants, and throwing stones or paper planes. Differences were small or absent in \textit{Shouting Man with Gun} and \textit{Solitary Confinement}. For arousal, effects of interaction were mixed, with lower values in \textit{Tunnel}, but with little or no change in other scenes. 

Taken together, these patterns indicate that affective interaction design consistently elevated Dominance across different contexts, allowing participants to feel more in control of the emotional experience rather than passively receiving it. In contrast, its impact on Valence and Arousal varied across scenes, suggesting that interaction does not simply amplify emotions uniformly, but rather shapes the user's experience in a context-dependent manner.

\subsection{Engagement Time in VR Scenes} \label{sec:time}

\begin{figure}[t]
    \centering
    \includegraphics[width=\linewidth]{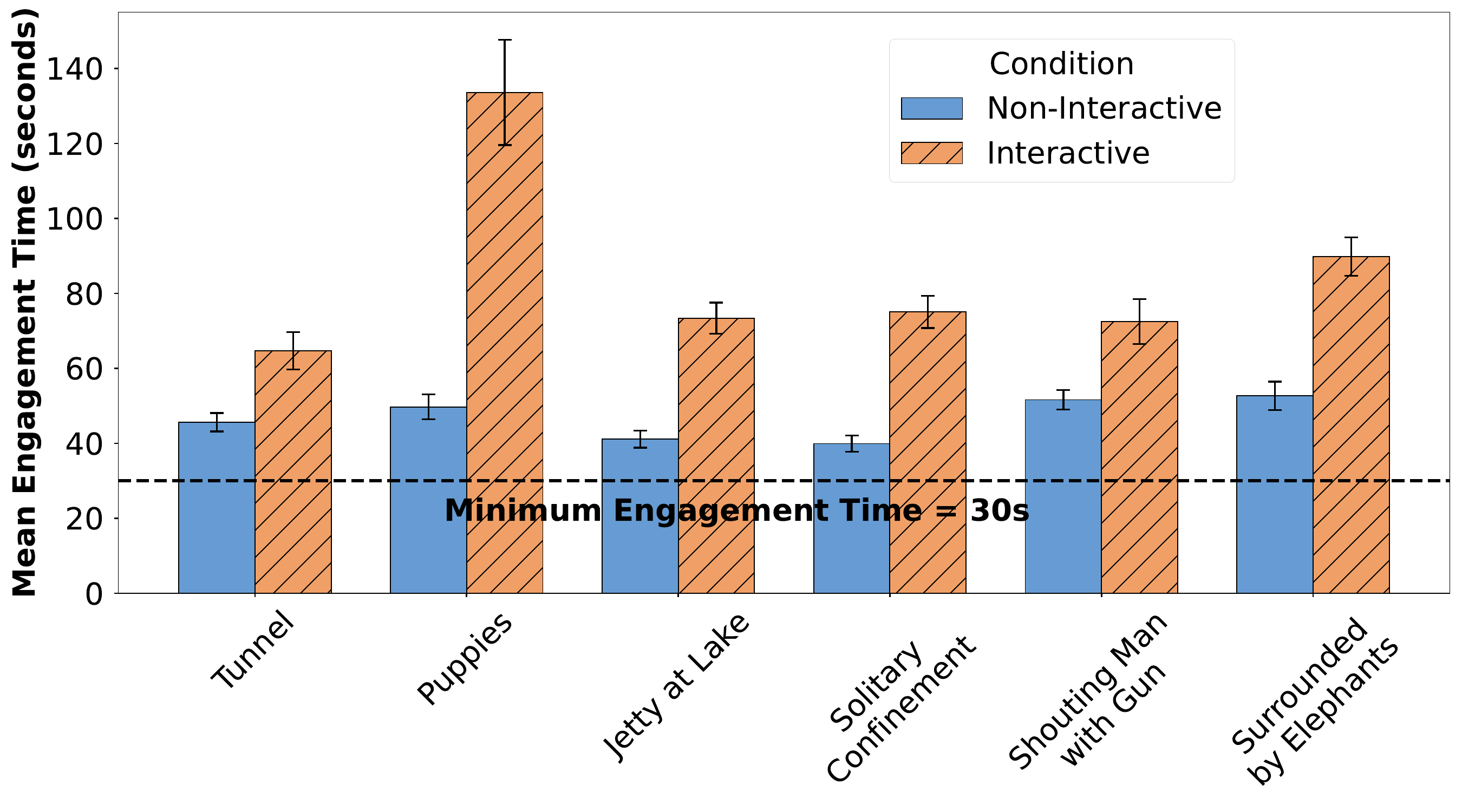}
    \caption{Mean engagement time across scenes for interactive and non-interactive conditions.}
    \label{fig:Engagement_Time}
    \Description{The y axis shows mean engagement time in seconds, and the x axis lists the six scenes: Tunnel, Puppies, Jetty at Lake, Solitary Confinement, Shouting Man with Gun, and Surrounded by Elephants. For each scene, two bars are shown: a solid blue bar for the non interactive condition and an orange hatched bar for the interactive condition, each with an error bar indicating variability. A horizontal reference line marks a minimum engagement time of 30 seconds. Engagement times are higher in the interactive condition for every scene, with the largest increase in Puppies.}
\end{figure}

Based on the SAM results, we assess user engagement time to understand how interaction influences the emotion-elicitation process. We compared the time participants spent in each scene under the interactive and non-interactive conditions (see \autoref{fig:Engagement_Time} for details). Engagement time (in seconds, error: $\pm 0.1$ s) was measured from the moment participants entered the scene until participants exited it. Following \citet{jiang2024immersive}'s approach, a minimum engagement time of 30 seconds was set to provide consistent exposure across scenes before participants could exit. The Shapiro–Wilk test indicated that engagement time distributions deviated from normality ($p < 0.05$). We thus used a linear mixed-effects model with participant as a random intercept to compare scenes across both conditions, revealing significant differences in mean engagement time.

The linear mixed-effects model revealed a significant main effect of interaction ($p < 0.001$), with participants spending longer in interactive than in non-interactive scenes. This effect was most pronounced in the \textit{Puppies} scene (interactive: $M = 133.58 \pm 89.92$ s, $Med = 113.87$ s; non-interactive: $M = 49.73 \pm 21.46$ s, $Med = 42.92$ s), where participants stayed more than twice as long on average when interaction was enabled. Substantial differences were also found in \textit{Jetty at Lake} ($M = 73.39$ s vs.\ $41.10$ s), \textit{Surrounded by Elephants} ($M = 89.77$ s vs.\ $52.66$ s), and \textit{Solitary Confinement} ($M = 75.05$ s vs.\ $39.93$ s), indicating that interaction extended engagement compared to the non-interactive scenes.

Overall, these results show that interaction significantly prolonged user exposure to the stimuli, suggesting that the opportunity to act encourages participants to sustain their engagement rather than exiting early. This indicates that affective interaction designs provide participants with greater opportunities to fully experience the intended emotional context within the virtual environment.

\subsection{Engagement with Virtual Environments}

To investigate how interaction shapes participants' spatial engagement and how these behavioral patterns correlate with emotional responses, we analyzed participants’ engagement with the virtual environments by comparing their virtual positions and orientations between the interactive and non-interactive conditions. For each scene, we identified the areas within the scenes where participants spent the most time. The analysis focused on the XZ plane, representing the top-down view of positions in the virtual space (in Unity, the Y dimension corresponds to height). \autoref{fig:CameraPoseCampairison} presents the sampled position data for each condition across scenes. By examining these spatial engagement patterns in combination with topic modeling results, we identified the following differences between the two conditions:
 
\begin{figure*}[t]
    \centering
    \includegraphics[width=1\linewidth]{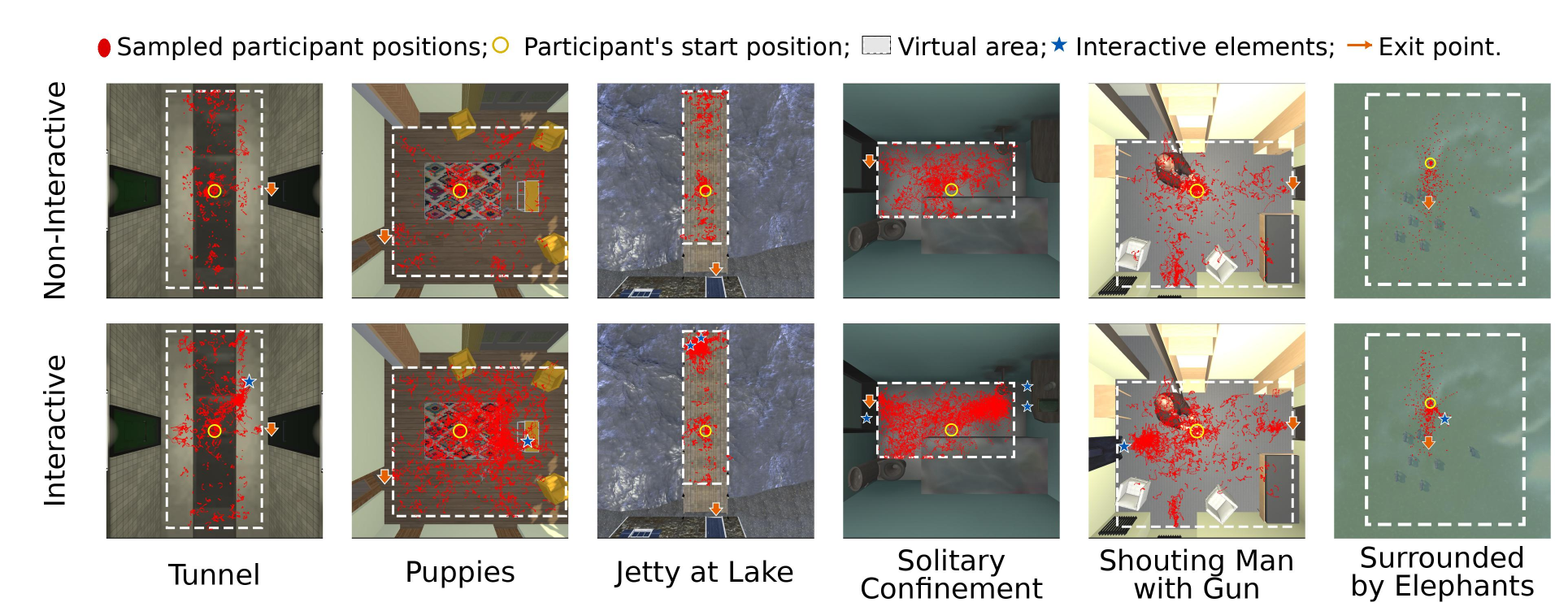}
    \caption{Spatial engagement patterns across the six VR scenes under non-interactive and interactive conditions. The positions are sampled every 0.1 seconds (sampling frequency = 10 Hz).}
    \label{fig:CameraPoseCampairison}
    \Description{The figure is arranged as two rows by six columns. The top row shows the non interactive condition and the bottom row shows the interactive condition. Each column corresponds to one scene (Tunnel, Puppies, Jetty at Lake, Solitary Confinement, Shouting Man with Gun, Surrounded by Elephants) and contains a top down view of the scene layout. Red dots mark sampled participant positions over time, producing dense clusters where participants spent more time. A yellow circle marks the participant’s start position, a dashed rectangle indicates the virtual area, blue star markers show the locations of interactive elements, and an orange arrow marks the exit point. Positions are sampled every 0.1 seconds (10 Hz), and the interactive row generally shows denser or more extended clusters around the interactive elements compared with the non interactive row.}
\end{figure*}

\paragraph{\textbf{Tunnel}} In the non-interactive condition, participants largely remained near the entrance of the corridor. With interaction enabled, participants advanced deeper into the tunnel after picking up the flashlight and explored the side doors, resulting in a more elongated spatial distribution. These behaviors were associated with higher dominance and lower arousal ratings in the interactive condition.

\paragraph{\textbf{Puppies}} Participants in the non-interactive condition mainly clustered near the entrance to observe the puppies. Under the interactive condition, they moved more widely across the room, engaging with the environment by petting the puppies or playing fetch using a tennis ball. This wider exploration corresponds to the higher valence, arousal, and dominance ratings.

\paragraph{\textbf{Jetty at Lake}} Movements in the interactive condition extended toward the end of the jetty, where stones and paper planes were located, creating clusters there. Compared to the limited exploration in the non-interactive condition, these behaviors were associated with higher valence and dominance as well as reduced arousal.

\paragraph{\textbf{Solitary Confinement}} Across both conditions, due to the nature of the scene, movement was highly restricted, with dense clusters near the spawning point. In the interactive condition, participants moved slightly toward the table and door to interact with the book, cup, or knocking on the door, resulting in a modest expansion of spatial distribution. These behaviors were accompanied with higher dominance and arousal, though valence remained low.

\paragraph{\textbf{Shouting Man with Gun}} In the non-interactive condition, participants often moved toward the window at the back of the room. By contrast, the interactive condition involved more frequent approaches to the riot shield along the left wall and to the gunman himself. This shows more active coping behaviors and corresponds to higher dominance, whereas both conditions showed low valence and high arousal.

\paragraph{\textbf{Surrounded by Elephants}} In the non-interactive condition, participants dispersed toward the periphery of the elephant herd, producing a wider spatial spread. With interaction, movements concentrated within the herd’s range, focusing on feeding bananas or touching elephants. These interactions yielded higher valence, lower arousal, and greater dominance, reflecting a more positive and controlled emotional experience.

Overall, the spatial engagement analysis shows that interactive elements encouraged participants to move beyond the spawning point and interact with salient features of the environment. This increased exploration was most consistently associated with higher dominance, while changes in valence and arousal varied depending on the emotional character of each scene. Collectively, these spatial patterns demonstrate that interactive elements function as attentional anchors, effectively guiding users' physical movement and focus toward key emotional features of the scene.

\subsection{Physiological Measures}

In addition to the before-mentioned measures, we also analyzed physiological responses recorded with the EmbracePlus wristband and processed the signals using \texttt{NeuroKit2} ~\cite{makowski2021neurokit2}. From the raw BVP data, we derived Heart Rate Variability (HRV) and Heart Rate (HR). From the raw EDA data, we extracted the phasic component (Skin Conductance Responses, SCRs), SCR frequency, and mean SCR amplitude as indicators of arousal~\cite{boucsein2012electrodermal, babaei2021critique}.

From BVP, we detected systolic peaks and derived inter-beat interval (IBI) as the basis for HRV and HR computation~\cite{mcduff2016cogcam}. Following the general HRV analysis procedure~\cite{mcduff2016cogcam, sarsenbayeva2019measuring}, we interpolated and resampled the IBI data at 4 Hz to align the signals in a uniform time interval, and removed long-term trends through detrending. We then obtained the HRV power spectrum from the detrended IBI data by applying a short-time Fourier transform (STFT) with a 64-sized sliding window.
Within each window, we calculated the High-Frequency (HF: 0.15–0.40 Hz) component of HRV and applied a natural log transformation to normalize the distribution. We also computed the mean HR within each segment. Higher HF HRV is typically interpreted as stronger parasympathetic activation, which shows lower physiological arousal~\cite{appelhans2006heart}. Similarly, lower HR also indicates reduced arousal~\cite{appelhans2006heart}.

\begin{figure}
        \centering
        \includegraphics[width=\linewidth]{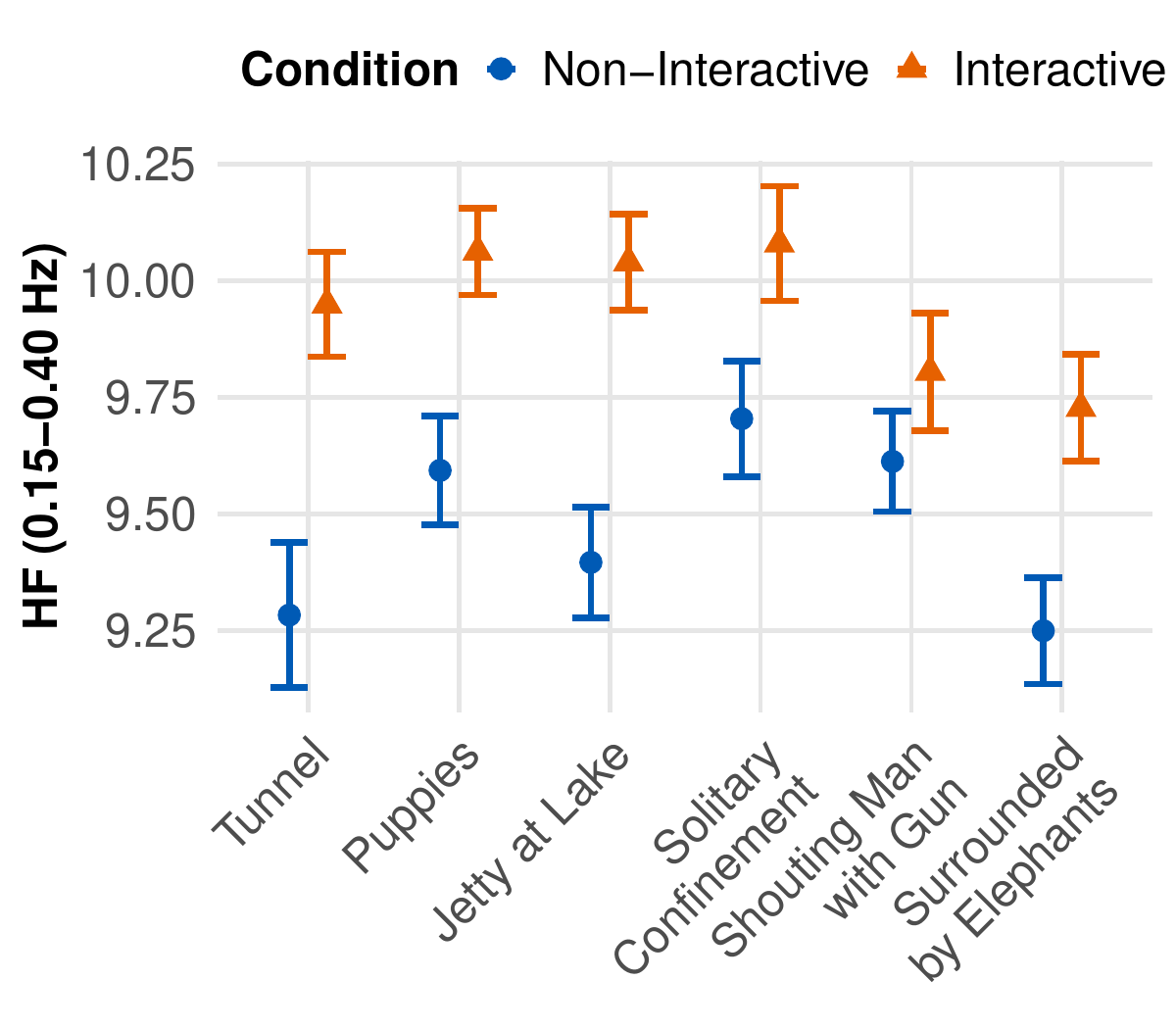}
        \caption{Fixed-effects estimates for high-frequency HRV (HF).}
        \label{fig:HF}
        \Description{The x axis lists the six scenes (Tunnel, Puppies, Jetty at Lake, Solitary Confinement, Shouting Man with Gun, Surrounded by Elephants). The y axis is HF (0.15–0.40 Hz), with values shown around 9.1 to 10.2. For each scene, two estimates are plotted: a blue circle for the non interactive condition and an orange triangle for the interactive condition. Vertical error bars show uncertainty around each estimate. Across all scenes, the interactive estimates are higher than the non interactive estimates.}
\end{figure}

\begin{figure}
        \centering
        \includegraphics[width=\linewidth]{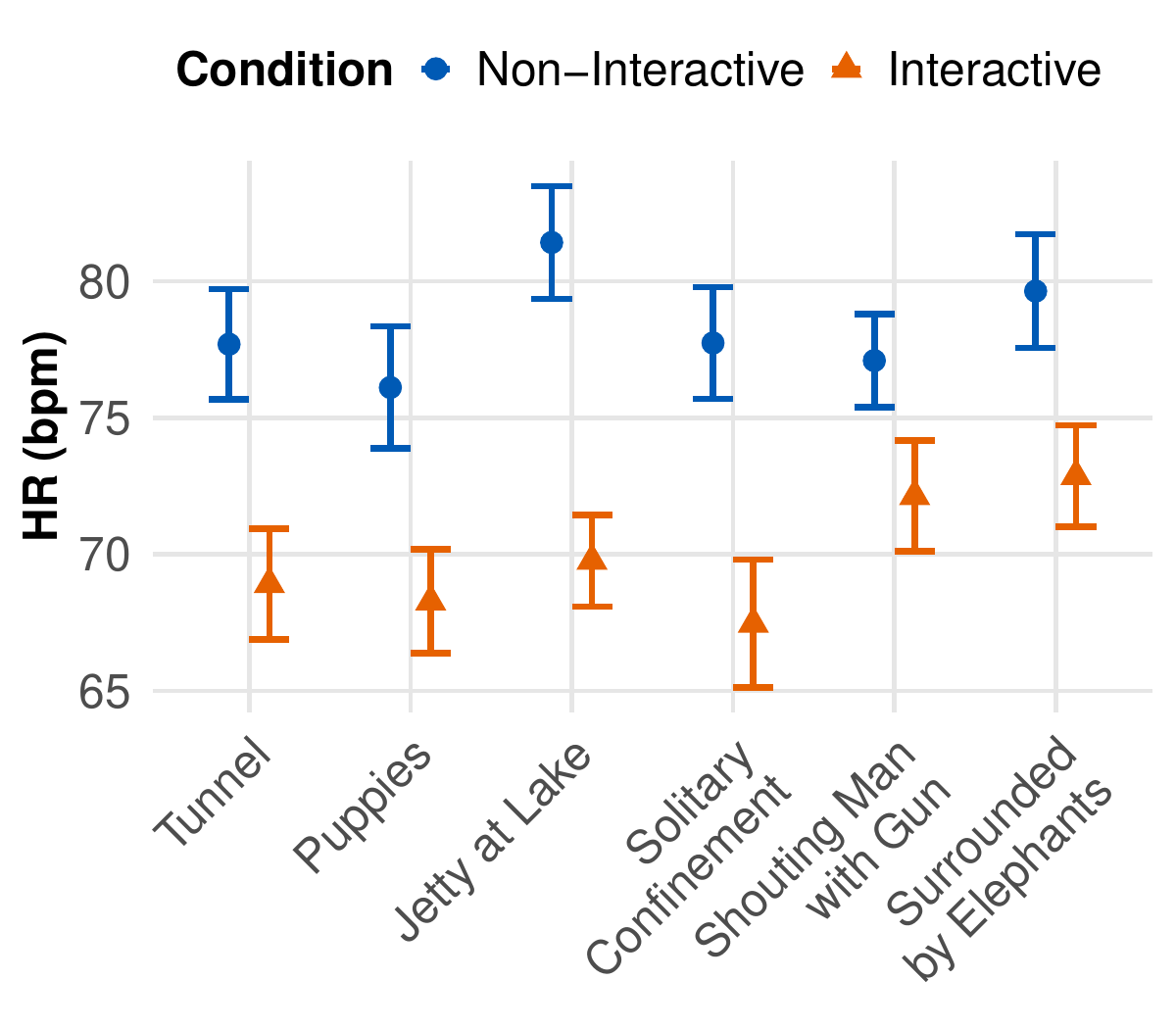}
        \caption{Fixed-effects estimates for heart rate (HR).}
        \label{fig:HR}
        \Description{The x axis lists the six scenes (Tunnel, Puppies, Jetty at Lake, Solitary Confinement, Shouting Man with Gun, Surrounded by Elephants). The y axis is heart rate in beats per minute. For each scene, two estimates are shown: a blue circle for the non interactive condition and an orange triangle for the interactive condition, each with vertical error bars indicating uncertainty. Across all scenes, the non interactive estimates are consistently higher than the interactive estimates, with the highest non interactive heart rate in Jetty at Lake and Surrounded by Elephants.}
\end{figure}

We applied linear mixed-effects models to the measures extracted from BVP signals, which revealed significant effects of interaction. Estimates are reported in \autoref{tab:physio_fixed_hf_hr} in \autoref{app:tables}, and  \autoref{fig:HF} and \ref{fig:HR}. HF HRV was significantly higher in the interactive condition (Estimate = 0.69, SE = 0.17, 95\% CI [0.36, 1.02]), while mean HR was significantly lower (Estimate = -9.25, SE = 2.82, 95\% CI [-14.80, -3.70]). Scene contrasts further showed increased HF HRV in \textit{Puppies} (Estimate = 0.31, SE = 0.12, 95\% CI [0.06, 0.55]), \textit{Solitary Confinement} (Estimate = 0.44, SE = 0.13, 95\% CI [0.19, 0.69]), and \textit{Shouting Man with Gun} (Estimate = 0.30, SE = 0.13, 95\% CI [0.05, 0.55]). Notably, in \textit{Shouting Man with Gun}, the increase in HF HRV associated with interaction was significantly attenuated (Estimate = -0.49, SE = 0.18, 95\% CI [-0.84, -0.14]), suggesting that interaction may have limited regulatory benefit in strongly negative contexts. This pattern is in line with the SAM ratings reported for these scenes.

We analyzed the phasic component of skin conductance responses (SCR). Following established procedures for EDA analysis ~\cite{boucsein2012electrodermal}, the raw signals were filtered and deconvolved to separate phasic activity from the tonic baseline. For each segment, we quantified two features: SCR count, representing the total number of responses, and mean SCR amplitude, representing the average response magnitude. Higher SCR count and amplitude are typically interpreted as stronger sympathetic activation, which reflects greater physiological arousal~\cite{boucsein2012electrodermal}.

\begin{figure}
    \centering
    \includegraphics[width=\linewidth]{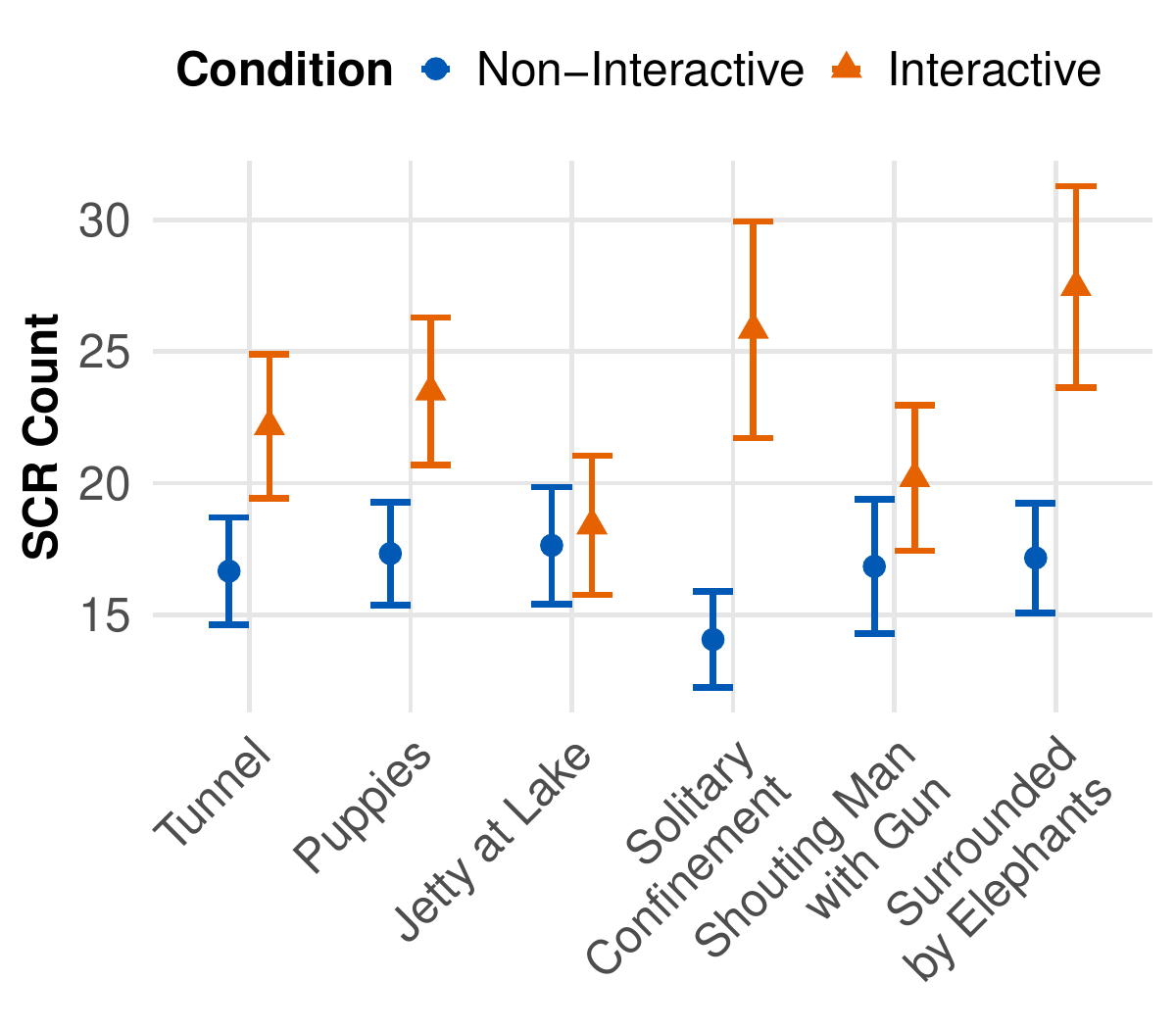}
    \caption{Fixed-effects estimates for SCR Count.}
    \Description{The x axis lists the six scenes (Tunnel, Puppies, Jetty at Lake, Solitary Confinement, Shouting Man with Gun, Surrounded by Elephants). The y axis shows SCR Count. For each scene, two estimates are plotted: a blue circle for the non interactive condition and an orange triangle for the interactive condition, with vertical error bars indicating uncertainty. Interactive estimates are higher than non interactive estimates for most scenes, especially Solitary Confinement and Surrounded by Elephants.}
    \label{fig:SCR_Count}
\end{figure}

\begin{figure}
    \centering
    \includegraphics[width=\linewidth]{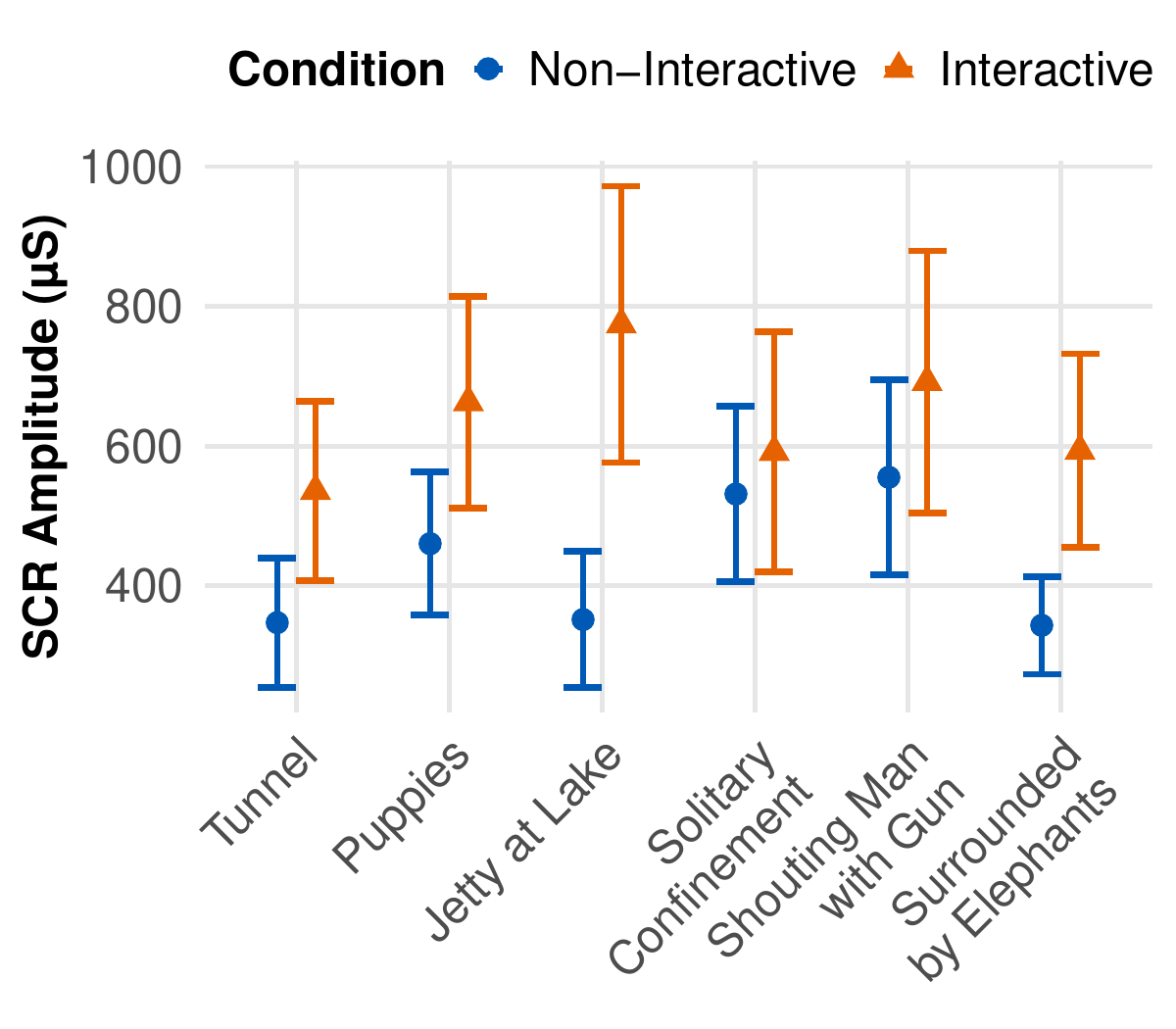}
    \caption{Fixed-effects estimates for SCR Amplitude.}
    \Description{The x axis lists the six scenes (Tunnel, Puppies, Jetty at Lake, Solitary Confinement, Shouting Man with Gun, Surrounded by Elephants). The y axis shows SCR amplitude in microSiemens. For each scene, two estimates are shown: a blue circle for the non interactive condition and an orange triangle for the interactive condition, with vertical error bars indicating uncertainty. Across all scenes, the interactive estimates are higher than the non interactive estimates, with the largest interactive amplitudes in Jetty at Lake and Shouting Man with Gun.}
    \label{fig:SCR_Amplitude}
\end{figure}

The linear mixed-effects models results for SCR count and SCR amplitude are shown in \autoref{tab:physio_fixed_scr} in \autoref{app:tables} and \autoref{fig:SCR_Count} and \ref{fig:SCR_Amplitude}. Although no statistically significant main effects of interaction or scene-level interactions were observed, consistent directional trends emerged that suggest increased sympathetic activation in the interactive condition. Specifically, for SCR count, marginal increases were observed across most scenes, with the largest differences in \textit{Tunnel}, \textit{Puppies}, \textit{Solitary Confinement}, and \textit{Surrounded by Elephants}. For SCR amplitude, a similar pattern was found, with stronger responses when interaction in the scenes was enabled, most prominently in \textit{Jetty at Lake} and \textit{Surrounded by Elephants}. These observations suggest that interaction elicited more frequent and stronger sympathetic responses.

In summary, the physiological results show that interaction led to significantly higher HF HRV and lower HR compared to the non-interactive scenes, while simultaneously eliciting consistent trends toward more frequent and stronger SCRs. This suggests that interaction creates a composite physiological state, characterized by overall relaxation interspersed with moments of high reactivity.

\subsection{Topic Modeling}
We further examined the emotion-elicitation process through participants’ qualitative descriptions collected in semi-structured interviews. Because emotions are inherently subjective, this analysis complements our quantitative results by capturing how participants themselves articulated their affective states. In particular, we focused on how these accounts differed between interactive and non-interactive scenes, providing richer insights into the role of interaction in shaping emotional experiences.

When analyzing the interview data, we applied topic modeling~\cite{jiang2024immersive, ma2023hello}. Transcripts were first preprocessed by retaining nouns and adjectives related to emotional qualities, while stop words and low-information terms (e.g., feeling, scene, little) were removed. All texts were lemmatized to normalize word variations (e.g., plurals to singular). We then used Latent Dirichlet Allocation (LDA)~\cite{blei2003latent} to extract three latent topics per scene, each represented by sets of co-occurring words that captured common themes in participants’ narratives~\cite{ma2023hello}. For interpretation, we manually examined the five most representative words for each topic and visualized overall word distributions with word clouds to illustrate how interactive elements shaped emotional expressions across different scenes.

\subsubsection{Topics: Tunnel}
\begin{figure*}[t]
    \centering
    \includegraphics[width=1\linewidth]{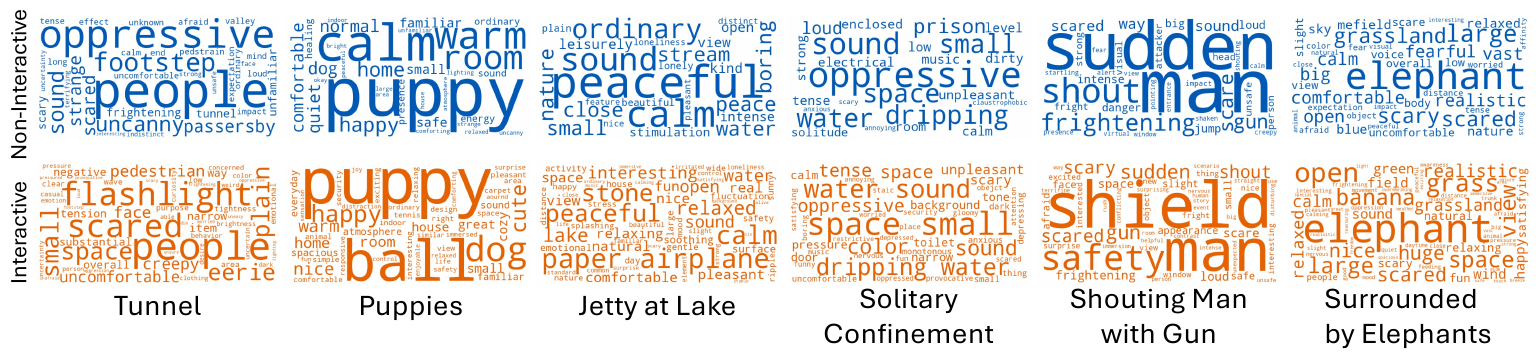}
    \caption{Word clouds of participants' descriptions of six VR scenes (top: Non-interactive, bottom: Interactive).}
    \label{fig:Topic}
    \Description{The figure has two rows and six columns. The top row shows word clouds for the non interactive condition and the bottom row shows word clouds for the interactive condition. Columns correspond to Tunnel, Puppies, Jetty at Lake, Solitary Confinement, Shouting Man with Gun, and Surrounded by Elephants. In each word cloud, larger words indicate terms mentioned more frequently. Common high frequency words include “puppy” and “calm” for Puppies, “peace” and “calm” for Jetty at Lake, “sudden” and “man” for Shouting Man with Gun, and “elephant” for Surrounded by Elephants; interactive versions additionally highlight words related to interaction and objects such as “flashlight,” “ball,” “shield,” and “safety.”}
\end{figure*}

The Tunnel scene placed participants in a long corridor with dim yellowish lighting and occasional pedestrians. In the non-interactive condition, it elicited mid valence ($M = 5.17 \pm 1.71$, $Med = 5.00$), mid arousal ($M$ = 5.02 ± 2.05, Med = 6.00), and neutral dominance ($M$ = 4.76 ± 2.08, Med = 5.00). Through tparticipants’ descriptions, as illustrated in \autoref{fig:Topic}, we observed mixed emotions described as ``oppressive'' (N = 3), ``uncanny'' (N = 2), ``scary'' (N = 2), ``strange'' (N = 2), and ``frightening'' (N = 2). LDA further highlighted themes of fear, oppression, and the unsettling presence of passersby. These patterns were often associated with the tunnel’s confined structure, as one participant noted: \textit{``There was an uncanny feeling, and the environment was dim, which made me feel a bit scared.''} (P67). Others emphasized the unsettling effect of pedestrians, for example: \textit{``The passersby made me feel a bit scared.'' (P55) and ``It felt frightening, and the people seemed strange.''} (P65).

In the \textit{interactive version}, when participants were given a flashlight, the scene instead elicited mid valence ($M$ = 5.14 ± 1.79, Med = 5.00), lower arousal ($M$ = 4.10 ± 2.06, Med = 4.00), and higher dominance ($M$ = 5.79 ± 2.19, Med = 6.00). Topic modeling revealed salient words such as ``people'' (N = 6), ``flashlight'' (N = 4), ``scared'' (N = 4), and ``creepy'' (N = 2). Topic modeling (LDA) indicated that while elements of eeriness remained, participants emphasized the flashlight as a source of agency and reassurance, reducing uncertainty about people’s behavior. For instance, one participant stated: \textit{``The confined space and low brightness of the colors made me concerned about people’s behavior. I wanted to leave. Using the flashlight to see clearly showed that people were just walking normally, and their casual clothing reduced the sense of pressure.''} (P9). Another participant explained: \textit{``The environment was dark, and using the flashlight to light up the face made it less frightening.''} (P17).

Overall, the Tunnel scene evoked unease through its narrow structure and the movement of pedestrians. In the non-interactive condition, participants emphasized feelings of fear, strangeness, and lack of control. In contrast, in the interactive condition, participants reported that the flashlight afforded agency, making the experience less threatening by reducing arousal and enhancing dominance.

\subsubsection{Topics: Puppies}

The \textit{Puppies} scene elicited emotions with high valence ($M$ = 7.12 ± 1.25, Med = 7.00), low arousal ($M$ = 4.33 ± 2.02, Med = 5.00), and high dominance ($M$ = 6.33 ± 1.48, Med = 7.00) in the non-interactive condition. As illustrated in \autoref{fig:Topic}, participants frequently described the experience as relaxing and delightful, often using words such as ``puppy'' (N = 8), ``warm'' (N = 4), ``room'' (N = 3), and ``happy'' (N = 3). LDA showed that these responses clustered around themes of familiarity, comfort, and calmness, with puppies and the indoor setting reinforcing feelings of safety and warmth: \textit{``The puppies made me happy, and the indoor environment felt very familiar and everyday''} (P67). 

When \textit{interaction was enabled}, participants could pet and play with the dogs, which heightened their affective engagement. Under this condition, the scene was rated with even higher valence ($M$ = 7.95 ± 1.17, Med = 8.00), slightly greater arousal ($M$ = 4.83 ± 2.62, Med = 5.00), and higher dominance ($M$ = 7.62 ± 1.43, Med = 8.00). Topic modeling highlighted terms such as ``puppy'' (N = 16), ``ball'' (N = 12), ``dog`` (N = 8), ``happy'' (N = 8), and ``cute'' (N = 6). LDA indicated themes of joy, playfulness, and comfort, emphasizing how active interaction with the puppies created both excitement and warmth. Participants highlighted the enjoyment of immersive engagement: \textit{``I enjoyed that when I threw the ball, the dogs brought it back.''} (P7). Others emphasized warmth and comfort: \textit{``Playing with the ball to tease the dog was very fun. The puppies were cute, and the tactile sensation felt nice.''} (P35). This emphasis on active interaction with the puppies also aligns with the observations in \autoref{sec:time}, where participants spent longer time in interactive scenes, suggesting that direct interaction with the puppies contributed to both longer engagement in the scene and stronger affective responses.

While both conditions elicited highly positive emotions, the interactive version strengthened these responses by introducing playfulness and tactile engagement, leading to greater immersion, joy, and a stronger sense of control.

\subsubsection{Topics: Jetty at Lake}

The \textit{Jetty at Lake} scene elicited emotions with high valence ($M$ = 6.86 ± 1.42, Med = 7.00), low arousal ($M$ = 4.33 ± 2.41, Med = 4.00), and high dominance ($M$ = 6.67 ± 1.88, Med = 7.00) in the non-interactive condition. Participants frequently associated the scene with calmness and serenity, using terms such as ``peaceful'' (N = 6), ``calm'' (N = 4), and ``sound'' (N = 3). LDA confirmed these impressions, highlighting themes of calmness, restorative nature, and quietness, with occasional mentions of loneliness or ordinariness. Several participants pointed to the natural setting as restorative: \textit{``It felt peaceful and close to nature.''} (P55); \textit{``The sound of water made me feel increasingly calm.''} (P52).

In the \textit{interactive version}, participants could throw paper airplanes and stones into the lake, which altered the quality of the experience. Under this condition, valence increased ($M$ = 7.48 ± 1.35, Med = 8.00), arousal slightly decreased ($M$ = 4.07 ± 2.42, Med = 4.00), and dominance was rated higher ($M$ = 7.26 ± 1.65, Med = 7.50). As we observed in \autoref{fig:Topic}, participants frequently mentioned ``paper airplane'' (N = 9), ``stone'' (N = 7), ``peaceful'' (N = 7), ``calm'' (N = 8), and ``relaxed'' (N = 4). LDA suggested that while the themes of peace and nature remained central, the addition of interactive elements introduced feelings of fun, relaxation, and control. Participants described them as enjoyable and soothing: \textit{``It felt open, comfortable, and interesting. The paper airplane and throwing stones made me feel calm''} (P6); \textit{``The paper airplane and stones helped me release stress, and the ripples from throwing them felt peaceful''} (P9).

Across conditions, participants consistently associated the \textit{Jetty at Lake} with relaxation. The interactive features deepened this effect by combining the natural scenery with simple actions, making the experience tranquil and engaging. The addition of paper airplanes and stones thus reinforced the calming qualities of the scene while enhancing participants’ sense of involvement and control.

\subsubsection{Topics: Solitary Confinement}
The \textit{Solitary Confinement} scene elicited emotions with low valence ($M$ = 3.74 ± 1.89, Med = 4.00), mid arousal ($M$ = 5.17 ± 2.12, Med = 5.00), and low dominance ($M$ = 3.69 ± 1.98, Med = 3.50) in the non-interactive condition. Participants frequently mentioned ``oppressive'' (N = 6), ``sound'' (N = 4), ``space'' (N = 3), ``small'' (N = 3), and ``dripping water'' (N = 3). LDA revealed themes of small space, disturbing sounds, and the prison setting. Participants often described the space as narrow and uncomfortable: \textit{``It felt oppressive, and the room was small.''} (P53). Others pointed to the audio design as particularly disturbing: \textit{``The dripping water sound was loud and annoying, making me feel anxious.''} (P79).

In the \textit{interactive condition}, where participants could interact with the cup, book, and door, the scene elicited slightly higher valence ($M$ = 4.52 ± 2.02, Med = 5.00), similar arousal ($M$ = 5.10 ± 2.09, Med = 5.00), and higher dominance ($M$ = 4.86 ± 2.53, Med = 5.00). Frequently mentioned words included ``space'' (N = 13), ``small'' (N = 11), ``sound'' (N = 10), ``dripping water'' (N = 5), and ``oppressive'' (N = 3). LDA indicated that although participants still described the space as confined and unpleasant, the possibility of interacting with the environment introduced a sense of relief or playfulness. For example, one participant noted: \textit{``It felt restrictive, the space was very small, and the dripping water sound made me feel tense and worried.''} (P36). Others highlighted playful aspects of interaction: \textit{``Throwing things at the door and hearing the sound felt provocative and satisfying.''} (P7).

Overall, while both conditions emphasized the oppressive qualities of the space and the disturbing soundscape, the interactive version allowed participants to vent their feelings and exert some control, making the experience slightly less negative.

\subsubsection{Topics: Shouting Man with Gun}

The \textit{Shouting Man with Gun} scene in non-interactive condition aimed to elicit emotions with low valence ($M$ = 4.33 ± 1.86, Med = 5.00), high arousal ($M$ = 6.81 ± 1.61, Med = 7.00), and low dominance ($M$ = 4.12 ± 1.98, Med = 4.00). Frequently mentioned words included ``man'' (N = 14), ``sudden'' (N = 14), ``shout'' (N = 13), ``frightening'' (N = 6), and ``gun'' (N = 4). LDA emphasized themes of sudden appearance, shouting, and danger, which reinforced feelings of fear and surprise. Participants described the scene as startling and frightening: \textit{``The man and his sudden shout were frightening''} (P63); \textit{``The view outside the window was calm, but the sudden entrance of the man was frightening''} (P72).

In the \textit{interactive condition}, ratings with higher valence ($M$ = 4.93 ± 2.15, Med = 5.00), identical arousal ($M$ = 6.83 ± 2.14, Med = 7.50), and higher dominance ($M$ = 4.50 ± 2.44, Med = 4.00). Participants frequently mentioned ``shield'' (N = 15), ``man'' (N = 13), ``safety'' (N = 8), ``sudden'' (N = 5), and ``gun'' (N = 4). LDA indicated that, while the man’s sudden entrance and shouting continued to evoke fear, the shield was perceived as protection that reduced helplessness. As one participant explained: \textit{``In the enclosed space, the man shouted, and his face and the way he pointed the gun at me made me feel scared. The shield, in contrast, gave me a sense of safety and immersion''} (P12). Another described: \textit{``The sudden event was unexpected and very frightening, but the shield gave me a slight sense of safety''} (P33).

In summary, the \textit{Shouting Man with Gun} scene elicited high-arousal, negative valence, dominated by fear and surprise. While both conditions produced similar responses, the interactive shield introduced a partial sense of safety, restoring a degree of agency.

\subsubsection{Topics: Surrounded by Elephants}
In the non-interactive condition, the \textit{Surrounded by Elephants} scene was rated with high valence ($M$ = 6.90 ± 1.76, Med = 7.00), high arousal ($M$ = 5.76 ± 2.21, Med = 6.00), and mid dominance ($M$ = 5.81 ± 2.30, Med = 6.00). Participants frequently mentioned ``elephant'' (N = 15), ``scared'' (N = 6), ``large'' (N = 4), ``grassland'' (N = 3), and ``comfortable'' (N = 2). LDA identified themes of openness, realism, and fear. One participant highlighted the vast natural setting: \textit{``The overall scene had an open view that matched my expectations.''} (P51). Others described mixed feelings: \textit{``It felt realistic, and it seemed like the elephant was going to attack me. I felt scared, fearful, and tense.''} (P57).

\textit{With interaction}, ratings remained highly positive, with higher valence ($M$ = 7.07 ± 1.44, Med = 7.00), slightly lower arousal ($M$ = 5.71 ± 2.43, Med = 6.50), and higher dominance ($M$ = 6.48 ± 2.03, Med = 7.00). Participants often mentioned ``elephant'' (N = 17), ``grassland'' (N = 8), ``banana'' (N = 4), ``large'' (N = 4), and ``relaxed'' (N = 3). LDA revealed themes of awe and nervousness, but also noted how feeding and touching the elephants provided moments of relief and enjoyment. As one participant observed: \textit{``The huge elephants created a slight sense of pressure and made me feel a bit nervous. The banana feeding interaction gave me a sense of awareness and made my mood feel calmer.''} (P9). P17 echoed this mix of fear and enjoyment: \textit{``The elephants were a bit scary because they were so big, but watching them eat bananas made me feel a bit more relaxed.''}

To summarize, the \textit{Surrounded by Elephants} scene elicited high-valence and high-arousal, with participants describing relaxation alongside moments of fear. According to our results, in the interactive condition, feeding and touching the elephants were often described as calming and enjoyable, relieving tension and fostering fun and involvement.

\subsubsection{Overview of topic modeling}
In summary, across all six scenes, the virtual environments effectively elicited the intended emotions, and our interactions with them further shaped participants’ responses. In the \textbf{non-interactive condition}, participants emphasized environmental atmosphere and passive reception, often reflecting a lack of control. In contrast, \textbf{the interactive condition} foregrounded agency and engagement. In negative scenes (e.g., \textit{Solitary Confinement} and \textit{Shouting Man with Gun}), our interaction designs provided avenues for coping and relief, whereas in positive or neutral scenes (e.g., \textit{Puppies} and \textit{Jetty at Lake}), they enhanced enjoyment and immersion. Overall, these results show that our interaction with the scenes did not simply intensify emotions but modulated them in context, highlighting the effectiveness of scene-tailored interaction in creating rich emotional experiences in VR.

\section{Discussion}

Next, we answer our research questions, position our work in existing work, and lay out implications and limitations. 

\subsection{RQ1: Does the added scene \textit{Surrounded by Elephants} reliably and effectively elicit High Arousal and High Valence emotions?} 

We extended the emotion elicitation dataset~\cite{jiang2024immersive} by adding a new VR scene, \textit{Surrounded by Elephants}, filling the gap in the HAHV quadrant. We validated this new VR scene through a user study, following the approach by \citet{jiang2024immersive}. Our findings reveal that compared to 360$\degree$ video, the VR scene elicited emotions with significantly higher valence and dominance. The higher valence ratings may suggest that participants experienced more positive emotions in the VR condition, consistent with prior research on positive emotion elicitation, which shows that an enhanced sense of presence in VR can strengthen affective responses~\cite{pavic2023feeling}. The higher dominance ratings further indicate that participants felt more engaged and were able to actively shape and control their experience through self-initiated behaviors in the VR environment, aligning with studies linking presence and control in VR to greater perceived dominance~\cite{riva2007affective, skarbez2020immersion}. In contrast, the difference in arousal was not significant, indicating that we found no difference between the VR scene and the 360$\degree$ video along this dimension. 

Overall, our scene demonstrated that the \textit{Surrounded by Elephants} VR scene can reliably and effectively elicit HAHV emotions, due to its immersive nature, as participants were able to \textit{e.g.}, approach or observe the elephants.

\subsection{RQ2: How does object-level interaction influence subjective and physiological measures of emotional response compared to a non-interactive baseline?}
We further extended the emotion elicitation dataset~\cite{jiang2024immersive} by adding interaction to the scenes.
Our findings demonstrate that, relative to the non-interactive baseline, \textit{object-level interaction} did not exhibit a uniform main effect on self-reported valence, arousal, or dominance, with differences appearing primarily as scene-dependent effects. However, the interaction patterns indicated that affective interaction design consistently elevated Dominance in specific contexts by allowing participants to feel more in control of the emotional experience rather than passively receiving it.

The interactive versions showed longer engagement time and broader spatial exploration, indicating that participants were more willing to stay and explore when given interaction options. This extended engagement provided more opportunities for emotional responses, reflected in higher valence and dominance ratings, as well as in physiological patterns of higher HF HRV and lower HR.

Furthermore, our findings align with prior work and show that \textit{object-level interaction} transforms passive experience into goal-directed exploration~\cite{christopoulos2018increasing}. Interactive objects provide clear action goals and feedback, prompting participants to actively perceive the virtual environment. Once participants know that interactive objects exist in the scene, they are more willing to remain within the scene and explore it. In low-valence or tense scenes, such as \textit{Tunnel}, \textit{Solitary Confinement}, and \textit{Shouting Man with Gun}, instrumental or defensive interaction offers ways to cope and regulate, reducing the tendency to exit quickly due to discomfort. Similarly, in scenes with high valence, such as \textit{Puppies}, \textit{Jetty at Lake}, and \textit{Surrounded by Elephants}, interaction seems to sustain engagement and enjoyment, thereby extending participation. 

Similarly, our findings demonstrate that \textit{object-level interaction} changes participants’ spatial exploration patterns in virtual scenes. Compared with the non-interactive version, spatial engagement shifts from simple surveying to goal-directed engagement with interactive objects. We observe that even when some objects are not configured as interactive, participants still attempt to engage with them, concentrating movement around potential interactive hotspots within a limited activity area. In the \textit{Puppies} scene, because the positions of the ball and the puppies can be changed, participants show broader spatial engagement with the environment. This goal-directed engagement has been shown to relate to emotional responses~\cite{fang2024exploring}: at the subjective level, it increases dominance, while the direction of valence and arousal depends on the affective character of the scene~\cite{mehrabian1974approach}; at the physiological level, it aligns with the overall pattern of higher HF HRV and lower HR~\cite{kim2018stress}, and increased SCR count and amplitude~\cite{critchley2002electrodermal}. Future work could extend this analysis by examining transient SCR fluctuations around key interaction events to capture finer-grained dynamics of physiological arousal~\cite{boucsein2012electrodermal}.

Overall, our findings suggest that interaction does not merely enhance emotional elicitation; by providing concrete goals, predictable feedback, and visible information, it reduces uncertainty and shapes the emotional experience in a controllable, context-sensitive way, rather than relying on passive exposure alone.

\subsection{RQ3: What is the relationship between subjective self-reports and physiological arousal in response to affective interactions in VR?}

In the interactive condition, across all scenes, we observed changes in physiological arousal that reflect two complementary dimensions. Higher HF HRV and lower HR indicate lower tonic arousal intensity, pointing to reduced physiological arousal~\cite{appelhans2006heart,kim2018stress}, while higher SCR count and amplitude reflected more frequent and stronger phasic activations, pointing to momentary sympathetic responses, e.g., HR accelerations, SCR responses~\cite{critchley2002electrodermal,boucsein2012electrodermal}. These physiological patterns consistently accompanied higher subjective valence and dominance, suggesting that participants felt more positive and in control, allowing them to maintain overall physiological relaxation interspersed with moments of high reactivity to interactive stimuli.

The relationship between subjective and physiological arousal showed dependence on the scene. For example, in the LALV scene \textit{Tunnel}, interaction reduced tension and uncertainty, leading to lower subjective arousal consistent with greater predictability and control. In \textit{Solitary Confinement}, interaction introduced executable goals and immediate feedback, shifting experience from passive low activation to object-directed engagement; subjective arousal and valence increased, while physiological measures indicated that this activation remained controllable rather than stressful. In the LAHV scenes, the effect depended on context: relaxation-oriented interaction reduced subjective arousal in \textit{Jetty at Lake}, whereas play-oriented interaction increased it in \textit{Puppies}. For high-arousal scenes, \textit{Surrounded by Elephants} (HAHV) showed only a modest decrease in subjective arousal, while \textit{Shouting Man with Gun} (HALV) yielded increased arousal despite weaker increases in HF HRV, suggesting vigilance combined with emerging control. 

In summary, interaction was associated with a consistent increase in valence and dominance, while physiological arousal showed a dual profile of tonic relaxation and phasic activation~\cite{appelhans2006heart, kim2018stress, critchley2002electrodermal}. Subjective arousal, in contrast, was more malleable, varying with scene context: decreasing in relaxation-oriented scenarios and rising under threat or active engagement~\cite{liszio2019interactive, valtchanov2010physiological, luong2022characterizing}. This partial dissociation between subjective and physiological arousal aligns with broader evidence that self-reports are shaped by cognitive appraisal of controllability and meaning~\cite{gross2015emotion}, whereas physiological measures capture the temporal dynamics of tonic versus phasic responses~\cite{babaei2021critique}, which may not always be directly mapped onto subjective experience~\cite{mauss2009measures, marin2020emotion}.

\subsection{Positioning within Existing Literature}

Compared to previous studies using the original 360$\degree$ videos~\cite{li2017public,schone2023library} and VR scenes~\cite{jiang2024immersive}, our findings were broadly consistent with prior work but revealed some systematic deviations. As shown in \autoref{fig:Emotional_Space_Comparison}, LALV and HALV scenes elicited higher valence at comparable arousal levels, while HAHV and LAHV scenes elicited higher arousal at comparable valence. These deviations may be explained by methodological factors. \citet{jiang2024immersive} conducted their study with Prolific participants completing VR experiences in different environments, where experimental control is limited, and variability can undermine consistency~\cite{mottelson2021conducting}. By contrast, our controlled laboratory setting ensured standardized equipment and procedures, conditions shown to enhance immersion and presence in VR~\cite{cummings2016immersive, slater2016enhancing}. To summarize, our findings align with prior work~\cite{jiang2024immersive}, but reveal systematic deviations in emotional responses, likely driven by the enhanced experimental control and immersion afforded by the laboratory setting compared to less controlled environments.

\subsection{Implications for Research and Design}

Our findings suggest that interaction in VR should be reconsidered not only as a method for eliciting emotions but also as a medium for modulating them. Specifically, we observed that interaction amplified positive affect in playful contexts, e.g., \textit{Puppies} and \textit{Surrounded by Elephants}, while reducing perceived threat and enhancing dominance in stressful scenes, e.g., \textit{Tunnel} or \textit{Solitary Confinement}. This indicates that the agency has the potential to function as a mechanism for emotional regulation in VR, reinforcing that immersive systems can shape coping and affective trajectories rather than simply intensifying responses~\cite{riva2019neuroscience, Liang_2025}. 
To facilitate such research, our validated dataset serves as a base for the HCI community to investigate how different object-level interactions influence emotional experiences in VR. Beyond the dataset, our findings offer a theoretical foundation for mental health research, specifically supporting a shift from passive exposure to active coping paradigms where affective interaction fosters a sense of control and safety~\cite{bell2020virtual, spytska2024use}.
Consequently, future work should extend beyond the valence–arousal model and develop measures that capture coping, relief, and sustained immersion. Such perspectives would position VR not only as an affective stimulus, but as a research tool for understanding and supporting emotion regulation across domains such as mental health~\cite{maples2017use}, training~\cite{kemeny2012contemplative}, and education~\cite{makransky2018structural}.

For design practice, our results highlight the importance of tailoring interaction to the specific affective qualities of a scene. In threatening contexts, giving users agency to act (\textit{e.g.}, controlling light in a dark tunnel) may alleviate distress and restore balance, while in positive contexts, playful or exploratory actions can heighten enjoyment and sustain engagement. These insights can be extended to diverse application areas. In VR game design, developers can adaptively adjust game difficulty or mechanics based on players' emotional state, providing empowering tools when anxiety is too high, or playful interactive objects when engagement drops~\cite{fairclough2009fundamentals}. In social VR, designing object-level interaction can provide shared focus and foster group collaboration and connection~\cite{scavarelli2021virtual}. In VR learning, playful interaction designs can promote active engagement and boost the sense of dominance, thereby enhancing learning outcomes~\cite{scavarelli2021virtual}.
However, effective design requires careful calibration. Prior studies on emotion elicitation in VR caution that poorly calibrated interaction risks overstimulation or heightened anxiety~\cite{riva2019neuroscience}, a concern our results corroborate. Thus, effective interaction design in VR requires balancing environmental atmosphere with user agency, ensuring that interaction supports adaptive regulation and meaningful emotional experiences.

\subsection{Limitations and Future Work}
Our study extends interaction research in VR by showing that object-level interaction can modulate emotions, amplifying positive affect while buffering negative intensity. At the same time, we identified several limitations that point to future directions. First, our focus was on object-level interactions; more complex forms, such as social interaction or dynamic environmental feedback, were not within our research scope. Some implemented actions may also resemble game mechanics, which could influence how users attribute and regulate their emotional responses across groups. Second, our measures relied on self-report (SAM) and basic physiological signals (EDA and BVP). While useful, these indicators are indirect and subject to substantial individual variability, limiting fine-grained interpretations of the regulation processes we observed.

Future work should broaden both the forms of interaction and the methods of assessment. Recent surveys highlight that multimodal affect recognition, integrating physiology with facial expressions, voice, and eye-tracking, provides a richer picture of emotional dynamics than single-modality approaches~\cite{abdullah2021multimodal}. Similarly, researchers argue that affective touch and embodied sensing are key to advancing emotion research in immersive contexts~\cite{zhang2025emotion}. These perspectives point beyond the valence–arousal–dominance model toward frameworks that capture coping, relief, and sustained engagement. Incorporating multimodal elicitation, including haptic feedback, may therefore enable more embodied emotional experiences in VR and deeper insights into how interaction might shape emotion elicitation and emotion regulation.

\section{Conclusion}
We extended a publicly available VR emotion elicitation dataset~\cite{jiang2024immersive} by creating and validating a new high-arousal, high-valence scene and systematically embedding object-level interaction across six VR scenes. Our comparison of interactive and non-interactive versions, analyzed through both self-report and physiological measures, shows that interaction operates as a mechanism of contextual emotion regulation: it supports coping in tense environments (e.g, with low valence) and enhances enjoyment in playful ones (e.g., with high valence) by balancing atmosphere, agency, and engagement. These findings provide empirical evidence that interaction shapes emotional responses in VR, advancing affective computing research, offering design insights for HCI applications in mental health, education, training, and interactive media, and also accelerating emotion research through immersive environments.

\begin{acks}
Weiwei Jiang is supported by the Natural Science Foundation of China under Grant 62402229 and the Startup Foundation for Introducing Talent of NUIST. Anusha Withana is a recipient of an Australian Research Council Discovery Early Career Award (DECRA) - DE200100479 funded by the Australian Government. 
Flora Salim would like to acknowledge the support of the ARC Centre of Excellence for Automated Decision Making and Society (CE200100005).
We thank our participants for their valuable time. We appreciate the members of the AID-LAB for assisting us in various ways.
\end{acks}

\bibliographystyle{ACM-Reference-Format}
\bibliography{bibliography}

\clearpage

\appendix
\section{Appendix}

\subsection{Questionnaire and Interview}

\subsubsection{Questionnaire} \label{app:pre-survey}

\begin{itemize}
    \item What is your gender identity? \\ $\Box$ Woman   $\Box$ Man   $\Box$ Non-binary/gender diverse   $\Box$ My gender identity is not listed 
    \item What is your current age (in years)? 
    \item How often do you use VR headsets? \\ $\Box$ Never   $\Box$ Daily   $\Box$ Weekly   $\Box$ Monthly
\end{itemize}

\subsubsection{Interview}

\begin{itemize}
    \item How would you rate your overall experience with the VR scenes? \\(Not enjoyable at all) 0   1   2   3   4   5 (Neutral)   6   7   8   9   10 (Extremely enjoyable)
    \item Did you experience any discomfort or dizziness during or after the VR experience?
    \item Which VR scene did you find the most emotionally impactful? \\ $\Box$ Tunnel $\Box$ Puppies $\Box$ Jetty at Lake \\ $\Box$ Solitary Confinement $\Box$ Shouting Man with Gun $\Box$ Surrounded by Elephants
    \item Why do you think that scene is the most emotionally impactful?
    \item Which VR scene did you find the least emotionally impactful? \\ $\Box$ Tunnel $\Box$ Puppies $\Box$ Jetty at Lake \\ $\Box$ Solitary Confinement $\Box$ Shouting Man with Gun $\Box$ Surrounded by Elephants
    \item Why do you think that scene is not as emotionally impactful as other scenes?
\end{itemize}

\subsubsection{Self-Assessment Manikin (SAM)}
\label{app:sam}
\leavevmode\par

\begin{figure}[!htb]
  \centering
  \includegraphics[width=\linewidth]{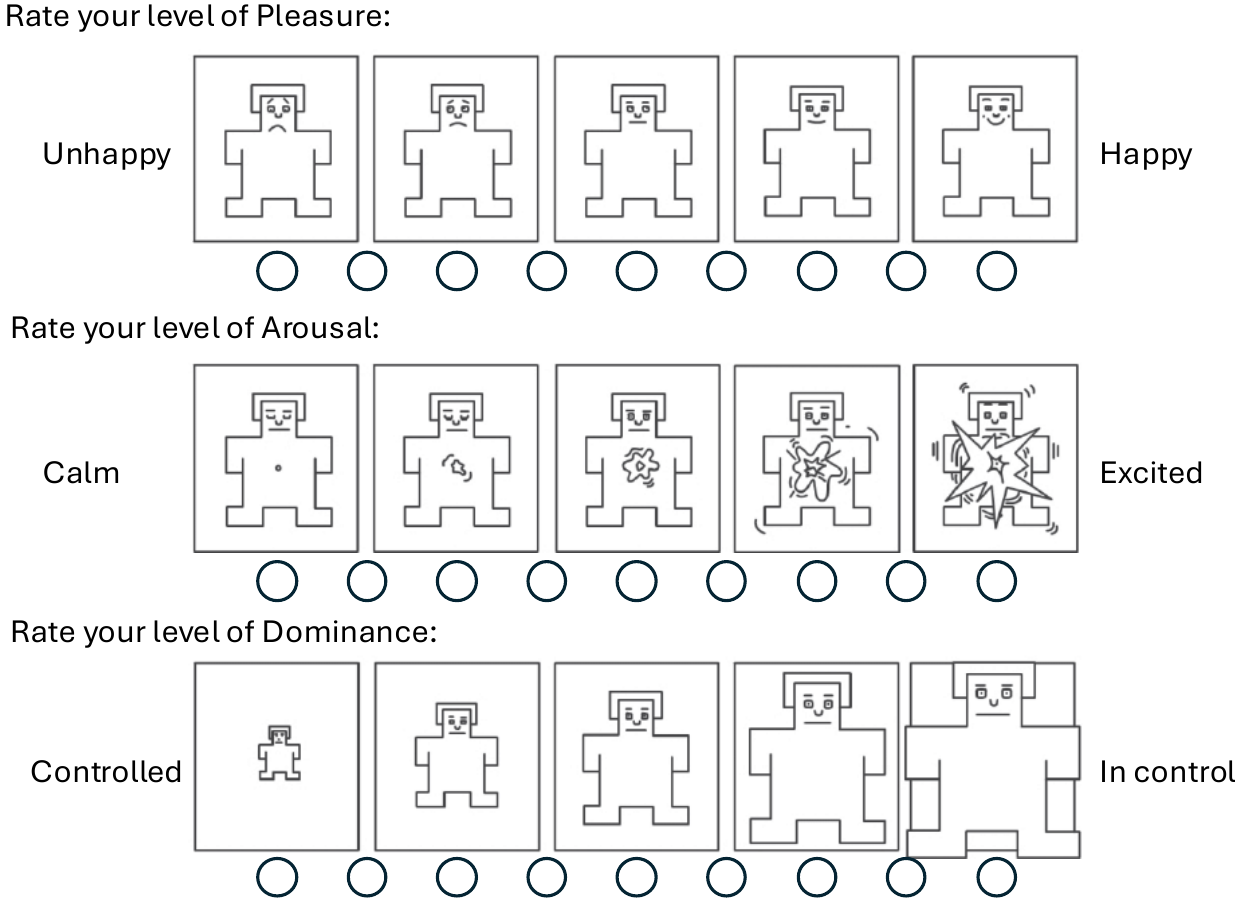}
  \caption{The self-assessment manikin (SAM) questionnaire.}
  \Description{The figure contains three horizontal rows. The first row asks participants to rate pleasure from “Unhappy” to “Happy” using a sequence of cartoon manikins whose facial expression becomes more positive. The second row asks participants to rate arousal from “Calm” to “Excited” using manikins that show increasing activation and movement. The third row asks participants to rate dominance from “Controlled” to “In control” using manikins that increase in size, indicating greater perceived control. Under each row, nine circular response options represent the selectable points on the scale.}
  \label{fig:sam}
\end{figure}

\newpage

\subsection{Interactive Objects}
\label{app:objects}
\begin{figure}[hbt!]
  \centering
  \setlength{\tabcolsep}{2pt}
  \captionsetup[sub]{labelformat=empty, skip=2pt, belowskip=4pt,  justification=centering}
  \begin{tabular}{ccc}
    \subcaptionbox{Flashlight\label{fig:tunnel}}[0.32\linewidth]{%
      \includegraphics[width=\linewidth]{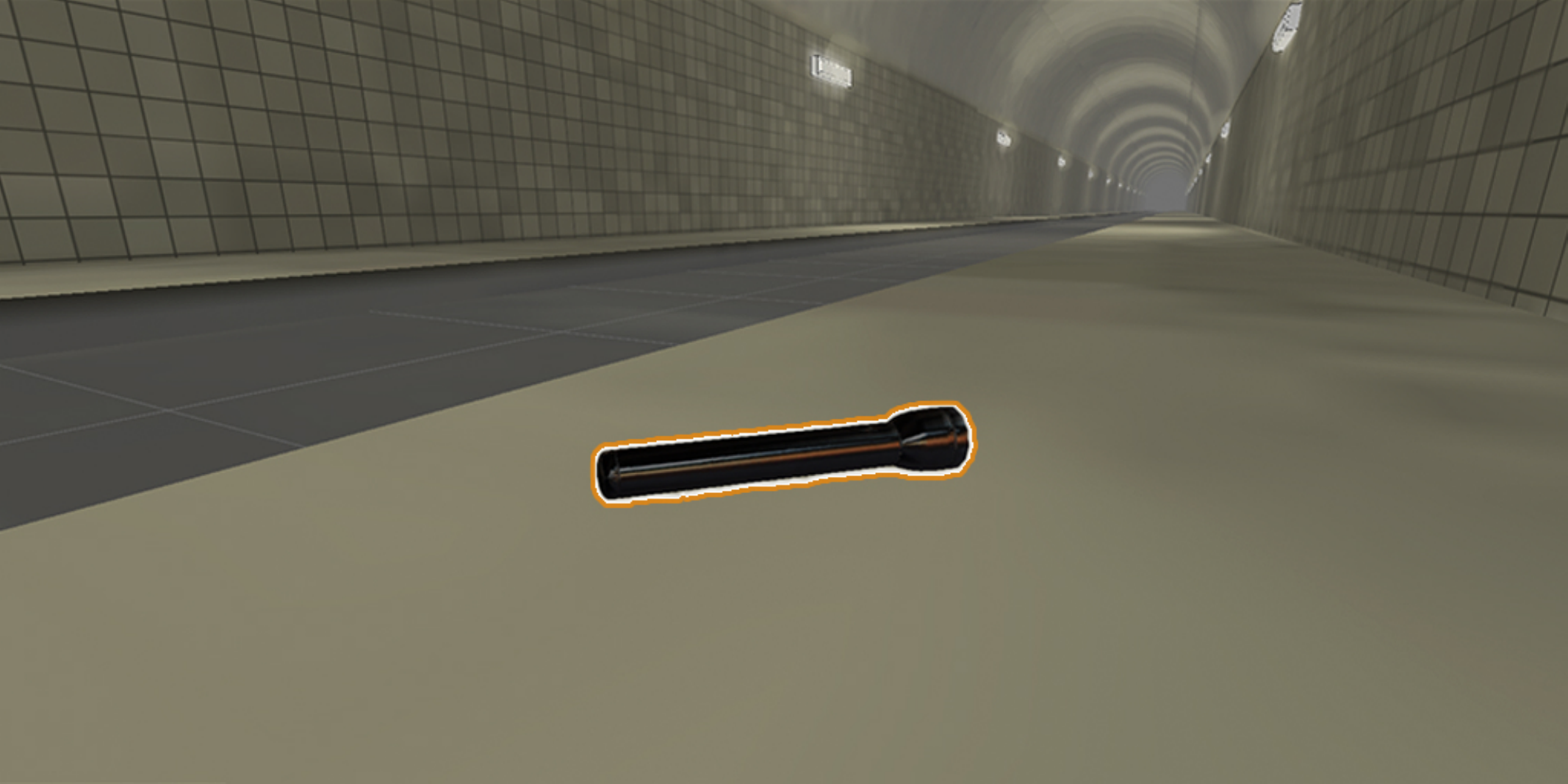}} &
    \subcaptionbox{Puppies and tennis ball\label{fig:puppies}}[0.32\linewidth]{%
      \includegraphics[width=\linewidth]{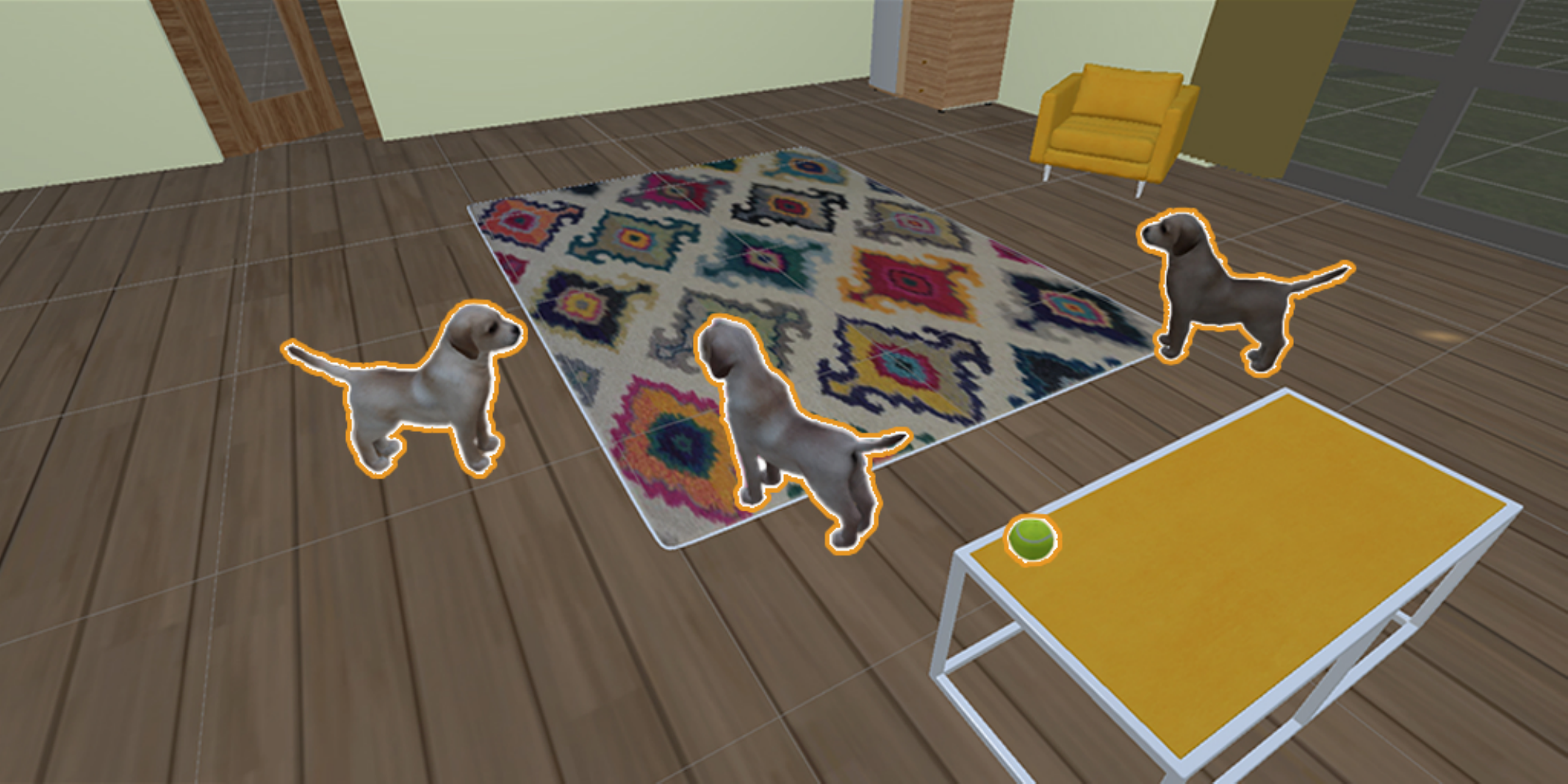}} &
    \subcaptionbox{Paper airplane and stones\label{fig:jetty}}[0.32\linewidth]{%
      \includegraphics[width=\linewidth]{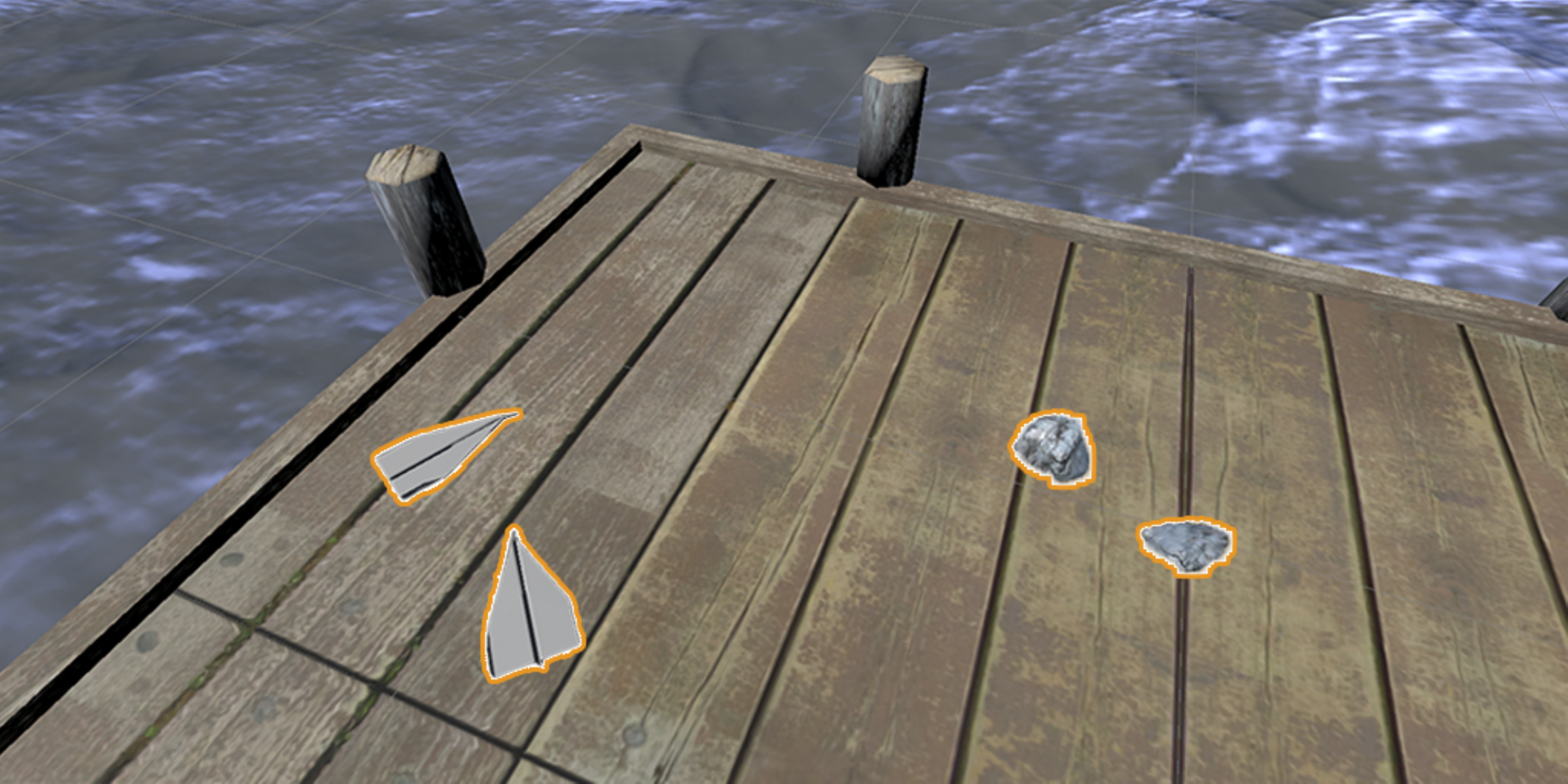}} \\[8pt] 
    \subcaptionbox{Book, metal cup and door\label{fig:solitary}}[0.32\linewidth]{%
      \includegraphics[width=\linewidth]{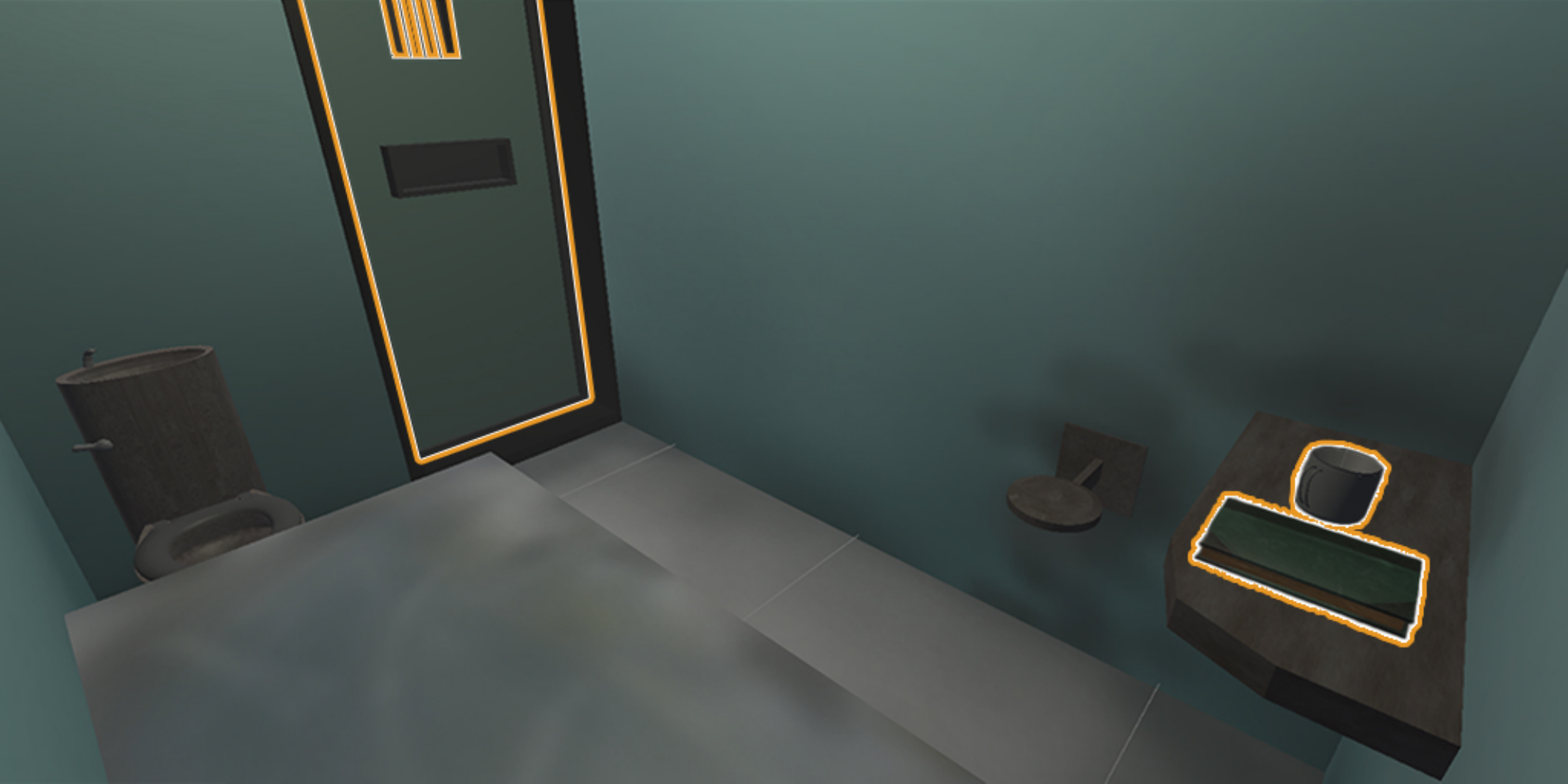}} &      
    \subcaptionbox{Metal riot shield\label{fig:gun}}[0.32\linewidth]{%
      \includegraphics[width=\linewidth]{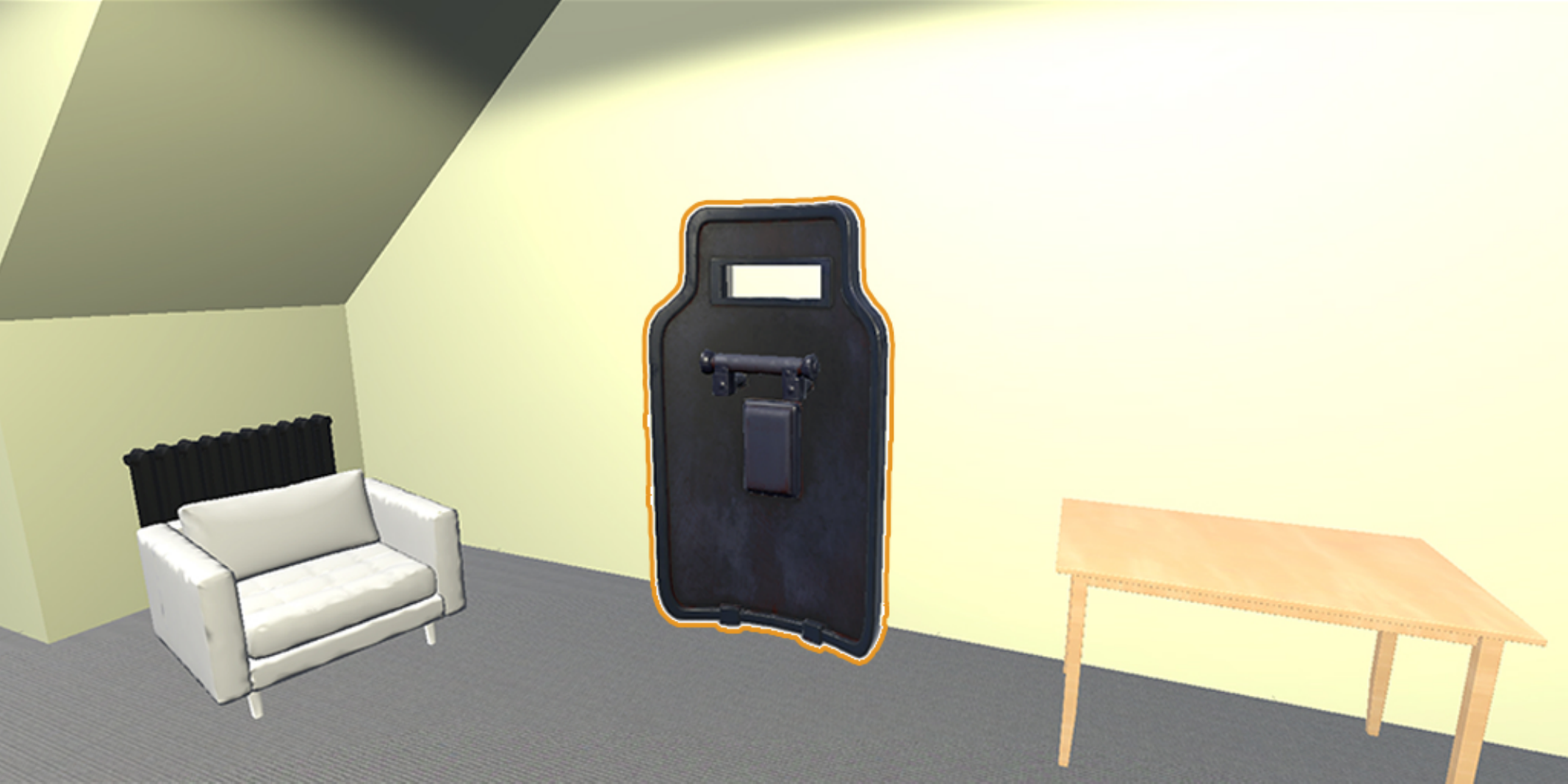}} &
    \subcaptionbox{Banana and elephants\label{fig:elephants}}[0.32\linewidth]{%
      \includegraphics[width=\linewidth]{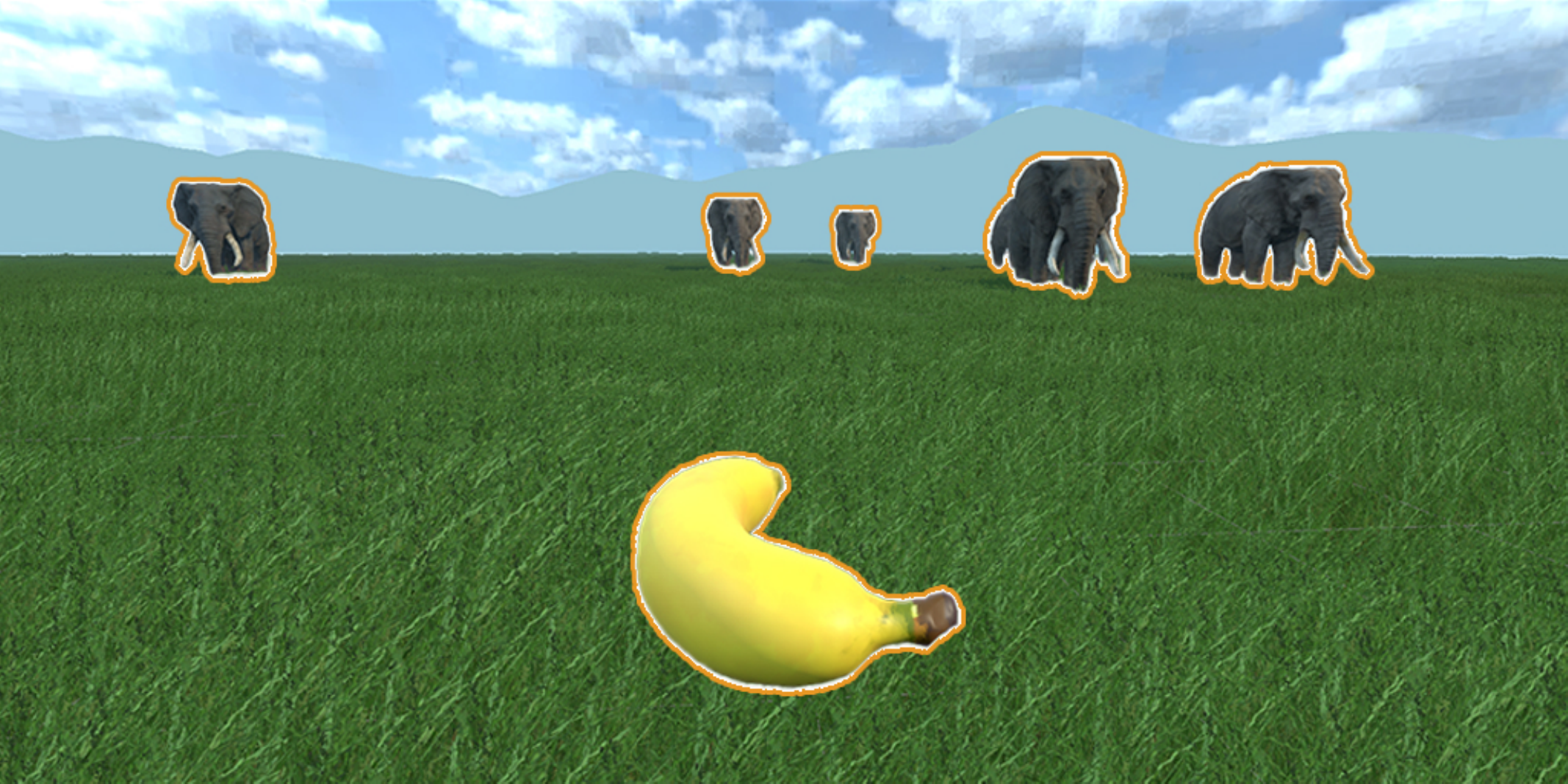}}    
  \end{tabular}
  \caption{Interactive objects with yellow outlines in the six VR scenes.}
  \Description{Six small scene screenshots are arranged in two rows of three, each highlighting interactive items with yellow outlines. The panels are labeled: Flashlight; Puppies and tennis ball; Paper airplane and stones; Book, metal cup and door; Metal riot shield; and Banana and elephants. The images show each object in its scene context, such as a flashlight on the tunnel floor, outlined puppies and a ball in a room, outlined paper airplane and stones on a jetty, outlined book and cup near a door in solitary confinement, an outlined riot shield in an indoor space, and an outlined banana on grass with elephants in the background.}
  \label{fig:objects}
\end{figure}

\subsection{Supplementary Tables}
\label{app:tables}

\begin{table}[h]
  \centering
  \caption{Demographic details and VR experience of participants in the Interactive and Non-Interactive conditions.}
  \label{tab:demographics}
  \begin{tabular}{llcc}
    \toprule
    \multirow{2}{*}{\textbf{Metric}} & \multirow{2}{*}{\textbf{Category}} & \multicolumn{2}{c}{\textbf{Condition}} \\
    \cmidrule(l){3-4}
     &  & \textbf{Non-Interactive} & \textbf{Interactive} \\
    \midrule
    \multirow{2}{*}{\textbf{Gender}} & Male & 21 & 21 \\
     & Female & 21 & 21 \\
    \midrule
    \multirow{2}{*}{\textbf{Age}} & Mean & 23.79 & 24.31 \\
     & SD & 3.21 & 2.87 \\
    \midrule
    \multirow{4}{*}{\textbf{VR Exp.}} & Never & 13 & 11 \\
     & Daily & 7 & 7 \\
     & Weekly & 10 & 11 \\
     & Monthly & 12 & 13 \\
    \bottomrule
  \end{tabular}
\end{table}


\begin{table}[h]
\centering
\caption{Mean and Median SAM ratings for the \textit{Surrounded by Elephants} VR scene.}
\label{tab:elephant_sam}
\begin{tabular}{lcccccc}
\toprule
{\textbf{Condition}} & \multicolumn{2}{c}{\textbf{Valence}} & \multicolumn{2}{c}{\textbf{Arousal}} & \multicolumn{2}{c}{\textbf{Dominance}} \\
\cmidrule(lr){2-3} \cmidrule(lr){4-5} \cmidrule(lr){6-7}
 & M & Med & M & Med & M & Med \\
\midrule
Previous Study~\cite{li2017public} & 5.94 & NR & 5.56 & NR & NR & NR \\
360$^{\circ}$ Video   & 5.42 & 5.50 & 5.67 & 6.00 & 5.08 & 4.50 \\
VR Scene    & 6.71 & 7.00 & 5.54 & 6.00 & 6.29 & 7.00 \\
\bottomrule
\multicolumn{7}{l}{\footnotesize NR = Not reported in the previous study.}\\
\end{tabular}
\end{table}

\newpage

\begin{table*}[h]
\centering
\caption{Fixed-effects estimates from linear mixed-effects models for high-frequency HRV (HF) and heart rate (HR). Each outcome was modelled as \textit{Outcome} $\sim$ Condition * Scene + (1|Participant). Values are reported as Estimate (SE) with 95\% CI.}
\label{tab:physio_fixed_hf_hr}
\begin{tabularx}{\linewidth}{Xcccccc}
\toprule
\multirow{2}{*}{\textbf{Parameter}} &
\multicolumn{3}{c}{\textbf{HF}} &
\multicolumn{3}{c}{\textbf{HR}} \\
\cmidrule(r){2-4}\cmidrule(l){5-7}
& Estimate & SE & 95\% CI & Estimate & SE & 95\% CI \\
\midrule
Intercept                                   & \textbf{9.29\textsuperscript{***}} & 0.12 & [9.06, 9.52] & \textbf{77.67\textsuperscript{***}} & 1.99 & [73.74, 81.59] \\
Interactive                                 & \textbf{0.69\textsuperscript{***}} & 0.17 & [0.36, 1.02] & \textbf{-9.25\textsuperscript{**}} & 2.82 & [-14.80, -3.70] \\
Puppies                                     & \textbf{0.31\textsuperscript{*}} & 0.12 & [0.06, 0.55] & -1.55 & 1.93 & [-5.36, 2.25] \\
Jetty at Lake                               & 0.13 & 0.13 & [-0.12, 0.38] & 3.55 & 1.99 & [-0.37, 7.47] \\
Solitary Confinement                        & \textbf{0.44\textsuperscript{***}} & 0.13 & [0.19, 0.69] & -0.32 & 1.96 & [-4.18, 3.54] \\
Shouting Man with Gun                       & \textbf{0.30\textsuperscript{*}} & 0.13 & [0.05, 0.55] & 0.09 & 1.98 & [-3.79, 3.98] \\
Surrounded by Elephants                     & -0.01 & 0.13 & [-0.26, 0.23] & 1.74 & 1.96 & [-2.11, 5.59] \\
Interactive $\times$ Puppies                & -0.22 & 0.18 & [-0.56, 0.13] & 1.46 & 2.75 & [-3.94, 6.86] \\
Interactive $\times$ Jetty at Lake          & -0.04 & 0.18 & [-0.39, 0.31] & -2.74 & 2.79 & [-8.22, 2.74] \\
Interactive $\times$ Solitary Confinement   & -0.34 & 0.18 & [-0.69, 0.01] & -0.39 & 2.78 & [-5.84, 5.07] \\
Interactive $\times$ Shouting Man with Gun  & \textbf{-0.49\textsuperscript{**}} & 0.18 & [-0.84, -0.14] & 3.87 & 2.79 & [-1.60, 9.35] \\
Interactive $\times$ Surrounded by Elephants& -0.24 & 0.18 & [-0.58, 0.11] & 2.71 & 2.75 & [-2.70, 8.12] \\
\bottomrule
\end{tabularx}\\
\vspace{0.5ex}
\raggedright\footnotesize
\textit{Notes.} Stars denote significance: \textsuperscript{*}\,$p<.05$, \textsuperscript{**}\,$p<.01$, \textsuperscript{***}\,$p<.001$. 
\end{table*}

\begin{table*}[h]
\centering
\caption{Fixed-effects estimates from linear mixed-effects models for SCR Count and SCR amplitude. Each outcome was modelled as \textit{Outcome} $\sim$ Condition * Scene + (1|Participant). Values are reported as Estimate (SE) with 95\% CI.}
\label{tab:physio_fixed_scr}
\begin{tabularx}{\linewidth}{Xcccccc}
\toprule
\multirow{2}{*}{\textbf{Parameter}} &
\multicolumn{3}{c}{\textbf{SCR Count}} &
\multicolumn{3}{c}{\textbf{SCR Amplitude}} \\
\cmidrule(r){2-4}\cmidrule(l){5-7}
& Estimate & SE & 95\% CI & Estimate & SE & 95\% CI \\
\midrule
Intercept                                   & \textbf{16.67\textsuperscript{***}} & 2.69 & [11.35, 21.98] & \textbf{347.31\textsuperscript{*}} & 136.68 & [77.05, 617.57] \\
Interactive                                 & 5.50  & 3.81 & [-2.01, 13.01] &  188.73 & 193.29 & [-193.48, 570.93] \\
Puppies                                     & 0.67  & 2.58 & [-4.40, 5.74]  &  113.03 & 102.73 & [-88.92, 314.98] \\
Jetty at Lake                               & 0.98  & 2.58 & [-4.10, 6.05]  &    4.39 & 102.73 & [-197.56, 206.34] \\
Solitary Confinement                        & -2.60 & 2.58 & [-7.67, 2.48]  &  184.08 & 102.73 & [-17.87, 386.03] \\
Shouting Man with Gun                       & 0.28  & 2.64 & [-4.90, 5.47]  &  187.77 & 105.04 & [-18.73, 394.27] \\
Surrounded by Elephants                     & 0.50  & 2.58 & [-4.57, 5.57]  &   -3.87 & 102.73 & [-205.82, 198.08] \\
Interactive $\times$ Puppies                & 0.77  & 3.66 & [-6.43, 7.96]  &    1.02 & 145.82 & [-285.65, 287.68] \\
Interactive $\times$ Jetty at Lake          & -4.74 & 3.65 & [-11.91, 2.43] &  233.83 & 145.28 & [-51.77, 519.43] \\
Interactive $\times$ Solitary Confinement   & 6.37  & 3.66 & [-0.83, 13.57] & -141.12 & 145.82 & [-427.78, 145.55] \\
Interactive $\times$ Shouting Man with Gun  & -1.57 & 3.72 & [-8.88, 5.73]  &  -47.73 & 147.99 & [-338.65, 243.19] \\
Interactive $\times$ Surrounded by Elephants& 4.79  & 3.65 & [-2.39, 11.96] &   61.25 & 145.28 & [-224.35, 346.85] \\
\bottomrule
\end{tabularx}\\

\vspace{0.5ex}
\raggedright\footnotesize
\textit{Notes.} Stars denote significance: \textsuperscript{*}\,$p<.05$, \textsuperscript{**}\,$p<.01$, \textsuperscript{***}\,$p<.001$. 
\end{table*}

\end{document}